\documentclass[acmsmall,screen,authorversion]{acmart}

\usepackage{tikz}
\usepackage{amsmath}
\usepackage{amsfonts}
\usepackage{listings}

\title{Algorithm 1019: A Task-based Multi-shift QR/QZ Algorithm with Aggressive Early Deflation}

\date{}

\author{Mirko Myllykoski}
\email{mirkom@cs.umu.se}
\orcid{0000-0002-3689-0899}
\affiliation{
  \institution{Department of Computing Science and HPC2N, Ume\r{a} University}
  \postcode{SE-901 87}
  \city{Ume\r{a}}
  \country{Sweden}
}

\thanks{This work is part of a project (NLAFET) that has received funding from the European Union's Horizon 2020 research and innovation programme under grant agreement No 671633. 
This work was supported by the Swedish strategic research programme eSSENCE and Swedish Research Council (VR) under Grant E0485301.}



\newcommand{\figurescale}{0.92}

\newtheorem{remark}{Remark}

\newcommand{\tcite}[1]{\citeauthor{#1} \citeyearpar{#1}}

\begin{abstract}
  The QR algorithm is one of the three phases in the process of computing the eigenvalues and the eigenvectors of a dense nonsymmetric matrix.
  This paper describes a task-based QR algorithm for reducing an upper Hessenberg matrix to real Schur form.
  The task-based algorithm also supports generalized eigenvalue problems (QZ algorithm) but this paper concentrates on the standard case.
  The task-based algorithm adopts previous algorithmic improvements, such as tightly-coupled multi-shifts and Aggressive Early Deflation (AED), and also incorporates several new ideas that significantly improve the performance.
  This includes, but is not limited to, the elimination of several synchronization points, the dynamic merging of previously separate computational steps, the shortening and the prioritization of the critical path, and experimental GPU support.
  The task-based implementation is demonstrated to be multiple times faster than multi-threaded LAPACK and ScaLAPACK in both single-node and multi-node configurations on two different machines based on Intel and AMD CPUs.
  The implementation is built on top of the StarPU runtime system and is part of the open-source StarNEig library.
\end{abstract}

\begin{document}

\setcopyright{acmlicensed}
\acmJournal{TOMS}
\acmYear{2021} \acmVolume{48} \acmNumber{1} \acmArticle{11} \acmMonth{12} \acmPrice{15.00}\acmDOI{10.1145/3495005}

\maketitle

\section{Introduction}

Given a matrix $A \in \mathbb{R}^{n \times n}$, the standard matrix eigenvalue problem consists of finding eigenvalues $\lambda_i \in \mathbb{C}$ and corresponding eigenvectors $x_i \in \mathbb{C}^n$, $x_i \neq 0$, such that
\begin{align}
 A x_i = \lambda_i x_i.
\end{align}
If the matrix $A$ is dense and nonsymmetric, then the eigenvalue problem is usually solved in three phases: (i) reduction to Hessenberg form, (ii) reduction to (real) Schur form, and (iii) computation of the eigenvectors.
The first phase reduces the dense matrix $A$ to upper Hessenberg form $H$ via an orthogonal similarity transformation and can be considered to be a preprocessing step.
The second step further reduces the upper Hessenberg matrix $H$ to (real) Schur form $S$ via the application of the QR algorithm.
After this step, the eigenvalues $\lambda_i$ can be determined from the diagonal blocks of the Schur form $S$.
Finally, the eigenvectors $x_i$ are computed from the Schur form $S$.
Note that although the Hessenberg reduction phase is generally considered to be the most time consuming phase, it can be accelerated with a GPU, see \tcite{tomov2009accelerating} and \tcite{myllykoski_2020}, and thus performed faster that the other two phases.
For a more detailed explanation of the three phases, see the classic textbook by \citeauthor{Golub1996} \citeyearpar{Golub1996}.

\paragraph{New task-based QR algorithm}

This paper focuses on the second phase and describes a task-based (see, e.g., \tcite{thibault}) QR algorithm that is implemented on top of the StarPU runtime system, see \tcite{starpu}.
The task-based algorithm also supports generalized eigenvalue problems (QZ algorithm) but this paper concentrates on the standard case.
The generalized case is relegated to a sequence of remarks highlighting the major differences.
The task-based algorithm adopts previous algorithmic improvements from the latest LAPACK, see \tcite{Byers07}, and ScaLAPACK, see \citeauthor{GraKagKreShao2015a} \citeyearpar{GraKagKreShao2015a, GraKagKreShao2015b}, implementations.
Specifically, the multi-shifts introduced by \tcite{10.1142/S0129053389000068}, tightly-coupled multi-shifts introduced by \tcite{Braman2002}, and Aggressive Early Deflation (AED) introduced by \tcite{Braman2002a}.
It also incorporates several new ideas that significantly improve the performance in both single-node and multi-node environments.
In particular, the new task-based algorithm introduces the following improvements:
\begin{enumerate}
 \item Elimination of several (global) synchronization points.
 In particular, the algorithm does not synchronize between bulge chasing and AED steps, or when a set of tightly-coupled bulges is moved across MPI process boundaries.
 \item Dynamic merging of previously separate computational steps.
 This leads to an increased computational resource utilization.
 \item Shortening and prioritization of the critical path. 
 This means that the critical path is completed sooner and lower-priority tasks are delayed until computational resources start becoming idle.
 \item Concurrent processing of several unreduced diagonal blocks.
 \item Adaptive approach for deciding when to perform a parallel AED.
 \item A common algorithm and implementation for both single-node and multi-node environments.
 \item GPU acceleration. 
\end{enumerate}

\paragraph{StarNEig library}

The new StarPU-based implementation is part of the open-source StarNEig library, see \citeauthor{myllykoski_2020} \citeyearpar{myllykoski_2020, ppam2019mm} and \tcite{starneig_website}.
This library aims to provide a complete task-based software stack for solving dense nonsymmetric standard and generalized eigenvalue problems.
Both single-node and multi-node environments are targeted and some components of the library support GPUs.
Currently, StarNEig implements the whole software stack for standard eigenvalue problems for single-node environments.
Support for multi-node environments is currently a work in progress. 
The Schur reduction phase is fully operational for both standard and generalized eigenvalue problems for both single-node and multi-node environments.
The missing software components are implemented as LAPACK and ScaLAPACK wrapper functions.
The development history of the StarNEig library, including the development history of the task-based QR/QZ algorithm, is documented in several NLAFET deliverable reports: \tcite{D25}, \tcite{D26}, \tcite{D27}, and \tcite{D65}.

\paragraph{Structure of the paper}

The rest of this paper is organized as follows: 
Section \ref{sec:qr} summarizes the modern QR algorithm, including the tightly-coupled multi-shifts and the AED, and discusses the related work.
Section \ref{sec:task_qr} describes the new task-based QR algorithm and its StarPU-based implementation.
Section \ref{seq:software} summarizes how the software is used in single-node and multi-node environments.
Section \ref{sec:results} demonstrates the performance of the StarPU-based implementation by comparing it against LAPACK and ScaLAPACK on Intel and AMD CPUs and Nvidia GPUs.
Note that MAGMA, see \tcite{magma}, and Elemental, see \tcite{elemental}, offload the Schur reduction phase to LAPACK and ScaLAPACK, respectively.
Other software packages, such as ELPA (\tcite{Marek_2014}), EigenEXA (\tcite{eigenexa}), PLASMA (\tcite{10.1007/s10766-016-0441-6}), and SLEPc (\tcite{slepc}), support only symmetric and/or sparse eigenvalue problems.
Finally, Section \ref{sec:conclusions} concludes the paper by summarizing the novel improvements and the major differences between LAPACK, ScaLAPACK and the new task-based algorithm.

\section{Modern QR algorithm and related work}\label{sec:qr}

The overall goal of the QR algorithm is to reduce an upper Hessenberg matrix
\begin{align}
 H = Q_1^T A Q_1 =
 \begin{bmatrix}
  h_{1,1} & h_{1,2} & \hdots    & h_{1,n} \\
  h_{2,1} & h_{2,2} & \hdots    & h_{2,n} \\
          & \ddots  & \ddots    & \vdots  \\
          &         & h_{n,n-1} & h_{n,n}
 \end{bmatrix} \in \mathbb{R}^{n \times n}
\label{eq:hessenberg_form}
\end{align}
to real Schur form
\begin{align}
 S = Q_2^T Q_1^T A \underbrace{Q_1 Q_2}_{=:\, Q} =
 \begin{bmatrix}
  S_{1,1} & S_{1,2} & \hdots & S_{1,m} \\
          & S_{2,2} & \hdots & S_{2,m} \\
          &         & \ddots & \vdots  \\
          &         &        & S_{m,m}
 \end{bmatrix} \in \mathbb{R}^{n \times n},
\label{eq:schur_form}
\end{align}
where $Q_1, Q_2 \in \mathbb{R}^{n \times n}$ are orthogonal matrices and
\begin{align}
 S_{1,1}, S_{2,2}, \dots, S_{m,m} \in  \mathbb{R}^{1 \times 1} \cup \{ X \in \mathbb{R}^{2 \times 2} : \lambda (X) \in \mathbb{C} \setminus \mathbb{R} \}.
\label{eq:2by2_blocks}
\end{align}
Above, $\lambda (X)$ denotes the set of all eigenvalues of a matrix $X$.
Now, since the matrix $Q$ is orthogonal and the matrix $S$ is upper quasi-triangular, we have
\begin{align}
 \lambda (A) = \lambda (H) = \lambda (S) = \bigcup_{i=1}^{m} \lambda (S_{i,i})
\end{align}
and the eigenvectors $x_i$ can be obtained by solving $(S - \lambda_i I) y_i = 0$ and backtransforming $x_i = Q y_i$.

Note that although the matrix $A$ is allowed to have complex eigenvalues and eigenvectors, all matrices involved have only real entries.
If the matrix $A$ has complex entries, then the involved computational phases are otherwise identical except that the matrix $S$ is complex and triangular, and all orthogonal similarity transformations are replaced with unitary similarity transformations.
Only the real case, i.e., the case where the matrix $A$ has only real entries but (possibly) complex eigenvalues and eigenvectors, is discussed in this paper.

\begin{remark}
In the generalized case, we are given a matrix pair $(A,B) \in \left(\mathbb{R}^{n \times n}\right)^2$ and we want to find generalized eigenvalues $\lambda_i \in \mathbb{C} \cup \{ \infty \}$ and corresponding generalized eigenvectors $x_i \in \mathbb{C}^{n \times n}$, $x_i \neq 0$, such that
\begin{align*}
 A x_i = \lambda_i B x_i.
\end{align*}
This is done by first reducing the matrix pair $(A,B)$ to Hessenberg-triangular form $(H,R) = Q_1 (A,B) Z_1^T,$ where $H$ is upper Hessenberg, $R$ is upper triangular and $Q_1, Z_1$ are orthogonal. 
The Hessenberg-triangular matrix pair $(H,R)$ is then further reduced to generalized Schur form
\begin{align*}
 (S,T) = Q_2^T (H,R) Z_2 =
 \left(
 \begin{bmatrix}
  S_{1,1} & S_{1,2} & \hdots & S_{1,m} \\
          & S_{2,2} & \hdots & S_{2,m} \\
          &         & \ddots & \vdots  \\
          &         &        & S_{m,m}
 \end{bmatrix},
 \begin{bmatrix}
  T_{1,1} & T_{1,2} & \hdots & T_{1,m} \\
          & T_{2,2} & \hdots & T_{2,m} \\
          &         & \ddots & \vdots  \\
          &         &        & T_{m,m}
 \end{bmatrix} \right) \in \left(\mathbb{R}^{n \times n}\right)^2,
\end{align*}
where $Q_2, Z_2$ are orthogonal and
\begin{align*}
 (S_{1,1}, T_{1,1}), \dots, (S_{m,m}, T_{m,m}) \in \left(\mathbb{R}^{1 \times 1}\right)^2  \cup \left\{ (X, Y) \in \left(\mathbb{R}^{2 \times 2}\right)^2 : \hat \lambda (X, Y) \in \mathbb{C} \setminus \mathbb{R} \right\}.
\end{align*}
Above, $\hat \lambda (X, Y)$ denotes the set of all generalized eigenvalues of a matrix pair $(X, Y)$.
In the end, we have
\begin{align*}
 \hat \lambda (A, B) = \hat \lambda (H, R) = \hat \lambda (S, T) = \bigcup_{i=1}^{m} \hat \lambda (S_{i,i}, T_{i,i}).
\end{align*}
The cases where $S_{i,i} \neq 0$ and $T_{i,i} = 0$ correspond to infinite eigenvalues.
\end{remark}

\subsection{Outline of the QR algorithm}

The QR algorithm is iterative.
More details will be covered in the following subsections but in summary, each iteration,
\begin{align}
\begin{split}
 H &\gets \hat Q^T_k \dots \hat Q^T_1 H \hat Q_1 \dots \hat Q_k, \\
 Q &\gets Q \hat Q_1 \dots \hat Q_k,
\end{split}
\end{align}
consists of a sequence of orthogonal transformations, $\hat Q_1 \dots \hat Q_k$, and produces an updated matrix $H$ that is otherwise of the form 
\begin{align}
\begin{bmatrix}
 H_{1,1} & H_{1,2} & \hdots & H_{1,\hat m} \\
         & H_{2,2} & \hdots & H_{2,\hat m} \\
         &         & \ddots & \vdots  \\
         &         &        & H_{\hat m,\hat m}
\end{bmatrix} \in \mathbb{R}^{n \times n},
\end{align}
but the condition
\begin{align}
 H_{i,i} \in  \mathbb{R}^{1 \times 1} \cup \{ X \in \mathbb{R}^{2 \times 2} : \lambda (X) \in \mathbb{C} \setminus \mathbb{R} \}.
\label{eq:2by2_cond}
\end{align}
does not necessarily hold for all $i = 1, \dots, \hat m$.
In particular, it is quite often the case that the diagonal blocks are larger than 2-by-2.
If the condition (\ref{eq:2by2_cond}) does not hold, we say that the matrix contains \emph{unreduced} blocks.
Note that this is also the case with the upper Hessenberg matrix (\ref{eq:hessenberg_form}) since the entire matrix consists of one unreduced block.

In most cases, the matrix contains a single unreduced block that gradually shrinks as the QR algorithm \emph{deflates} 1-by-1 and 2-by-2 diagonal blocks from the lower-right corner of the unreduced block.
However, an unreduced block can decouple into several independent unreduced blocks.
This can happen either because a sub-diagonal entry is flushed to zero or because the QR algorithm concludes that a sub-diagonal entry is so small that it can be set to zero without introducing significant perturbations (a.k.a \emph{vigilant deflation}).  
If an unreduced block decouples in this manner, then the QR algorithm is applied recursively to each one of the newly-created unreduced blocks.
The QR algorithm repeats until a matrix that does not contain any unreduced blocks is produced or an iteration limit is reached.

\begin{remark}
The QZ algorithm is iterative and very similar to the QR algorithm:
\begin{align*}
 (H, R) &\gets \hat Q^T_k \dots \hat Q^T_1 (H, R) \hat Z_1 \dots \hat Z_k, \\
 Q &\gets Q \hat Q_1 \dots \hat Q_k, \\
 Z &\gets Z \hat Z_1 \dots \hat Z_k.
\end{align*}
Note that the standard convention used by LAPACK is to classify a real eigenvalue as infinite if $|T_{i,i}| \leq u \|T\|_F$, where $\|\cdot\|_F$ is the Frobenius norm and $u$ is the unit roundoff.
In that case, $T_{i,i}$ is set to zero.
By default, the task-based algorithm uses the same convention.
\end{remark}

\subsection{Shifts and bulge chasing}

\begin{figure}[h]
 \centering
 \includegraphics[scale=\figurescale]{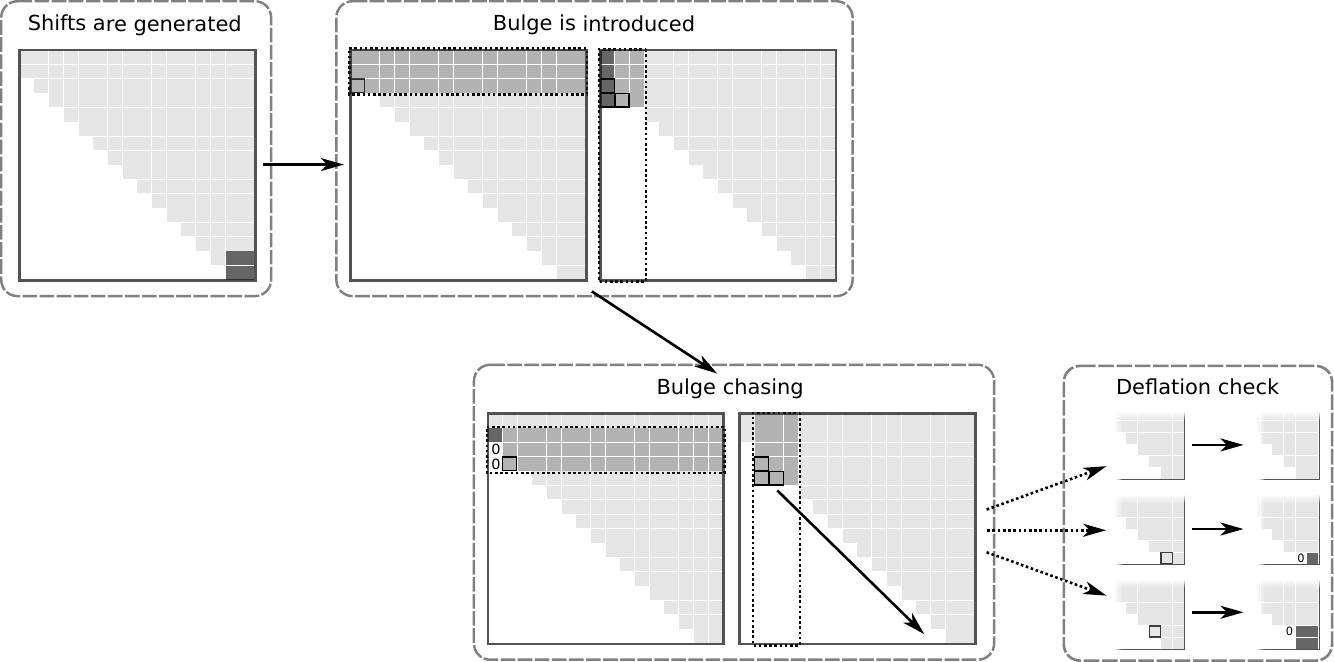}
 \caption{An illustration of the first iteration of the double-implicit-shift QR algorithm: generation of the shifts, introduction of the bulge, bulge chasing, and deflation check (from the top: no deflation, a 1-by-1 block / real eigenvalue deflated, and a 2-by-2 block / complex conjugate pair of eigenvalues deflated).}
 \label{fig:double_shift}
\end{figure}

Before discussing the modern incarnation of the QR algorithm, we will first remind the reader about the textbook double-implicit-shift QR algorithm, see \tcite{Golub1996}.
As illustrated in Fig. \ref{fig:double_shift}, each iteration begins with the computation of a pair of \emph{shifts}, $(\hat \lambda_1, \hat \lambda_2)$, by solving a small eigenvalue problem involving a 2-by-2 sub-matrix in the lower-right corner of an unreduced block $\hat H \in \mathbb{R}^{m \times m}$.
We then compute a vector $v = (\hat H - \hat \lambda_1 I)(\hat H - \hat \lambda_2 I)e_1$, generate a 3-by-3 Householder reflector\footnote{By a 3-by-3 Householder reflector, we mean an identity matrix $I \in \mathbb{R}^{m \times m}$ with a Householder reflector $X \in \mathbb{R}^{3 \times 3}$ embedded in it as a diagonal block.} $\hat Q_1$ such that $\hat Q_1^T v$ is a multiple of $e_1 = [1 \; 0 \; \dots \; 0]^T$, and apply $\hat Q_1$ to $\hat H$ from both sides: $\hat H \gets \hat Q_1^T \hat H \hat Q_1$.

The transformation produces a small 2-by-2 \emph{bulge} to the upper-left corner of $\hat H$ (see Fig. \ref{fig:double_shift}).
The unreduced block $\hat H$ is then returned back to upper Hessenberg form via a sequence of overlapping 3-by-3 Householder reflectors, $\hat Q_2 \dots \hat Q_k$, in a process called \emph{bulge chasing}.
Each iteration ends with the investigation of the sub-diagonal entries in the lower-right corner of $\hat H$.
If a sub-diagonal entry is found to be small enough, then it is set to zero and either a 1-by-1 or 2-by-2 diagonal block is decoupled from $\hat H$, thus deflating the corresponding eigenvalue or the corresponding complex conjugate pair of eigenvalues, respectively.
The implicit Q theorem (see \tcite{10.1093/comjnl/4.3.265}) tells us that the product matrix $\hat Q^T_k \dots \hat Q^T_1 H \hat Q_1 \dots \hat Q_k$ and the shifted matrix $(\hat H - \hat \lambda_1 I)(\hat H - \hat \lambda_2 I)$ are closely connected to each other (see \tcite{Golub1996}). 

\begin{remark}
In the generalized case, we compute a pair of shifts, $(\hat \lambda_1, \hat \lambda_2)$, by solving a small generalized eigenvalue problem involving a 2-by-2 sub-matrix pair in the lower-right corner of an unreduced block-pair $(\hat H, \hat T) \in \left(\mathbb{R}^{m \times m}\right)^2$.
We then compute a vector $v = (\hat H \hat T^{-1} - \hat \lambda_1 I)(\hat H \hat T^{-1} - \hat \lambda_2 I) e_1$, generate a 3-by-3 Householder reflector $\hat Q_1$ such that $\hat Q_1^T v$ is a multiple of $e_1 = [1 \; 0 \; \dots \; 0]^T$, and apply $\hat Q_1$ to $(\hat H, \hat T)$ from the left: $(\hat H, \hat T) \gets \hat Q_1 ^T(\hat H, \hat T)$.
The resulting matrix pair is then returned to Hessenberg-triangular form via a overlapping sequence of 3-by-3 Householder reflectors from the left and 3-by-3 ''right-hand side`` Householder reflectors from the right, see \tcite{Adlerborn2014} and \tcite{10.1137/S089547989122377X}.
In addition to the 2-by-2 bulge being chased down the diagonal of $\hat H$, a 1-by-1 bulge is simultaneously chased down the diagonal of $\hat T$.
\end{remark}

\subsection{Multi-shifts and BLAS-3 performance}\label{subsec:multishifts}

The modern incarnation of the QR algorithm improves upon the double-implicit-shift QR algorithm by introducing several algorithmic advances, such as the tightly-coupled multi-shifts and the AED.
We will first discuss the tightly-coupled multi-shifts (see \tcite{Braman2002}) that are used to improve the arithmetical intensity, and thus the performance, of the bulge chasing step.

\begin{figure}[h]
 \centering
 \includegraphics[scale=\figurescale]{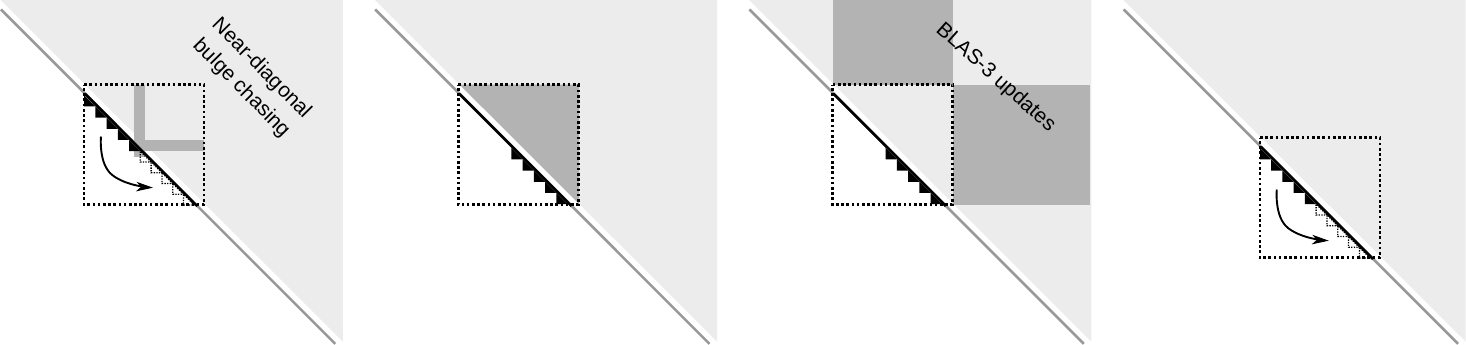}
 \caption{An illustration of near-diagonal bulge chasing and the related off-diagonal updates.}
 \label{fig:multi_shift}
\end{figure}

Instead of introducing just a single bulge and chasing it down the diagonal, the multi-shift QR algorithm introduces a \emph{set of tightly-coupled bulges}.
This is done by introducing the bulges one by one\footnote{The source of the necessary shifts is discussed in the Subsection \ref{subsec:aed}.} and then chasing them down the diagonal just enough so that the next bulge can be introduced. 
The bulges are then chased down the diagonal as a single unit, in a pipelined manner, as illustrated in Fig. \ref{fig:multi_shift}.
More specifically, the bulge chasing is initially performed only near the diagonal, inside a small \emph{diagonal window}.
The related 3-by-3 Householder reflectors are simultaneously accumulated into an \emph{accumulator matrix} and only later applied to the relevant off-diagonal regions using matrix-matrix multiplications (BLAS-3).
The bulge chasing step thus consists of a \emph{chain of overlapping bulge chasing windows} and the corresponding accumulated left-hand and right-hand side \emph{updates}.
The end result is a significantly improved arithmetical intensity compared to the double-implicit-shift bulge chasing step.
More advanced techniques, such as optimally packed chains\footnote{The word chain is used differently in this content as it refers to a chain of bulges instead of a chain of overlapping bulge chasing windows.} of bulges, see \tcite{10.1145/2559986} and \tcite{steel2020multishift}, can used to further improve the performance.
However, this advanced technique is not discussed in this paper as it is not yet implemented in any of the existing libraries.

\subsubsection{LAPACK implementation}

The LAPACK implementation chases all bulges together using a single bulge chasing window chain.
Since the off-diagonal updates are offloaded to BLAS-3 routines, the LAPACK implementation can be parallelized by linking LAPACK to a parallel BLAS library.
This means that each individual off-diagonal BLAS-3 update is performed in parallel.
As demonstrated in \tcite{D26}, this approach does not scale past a few cores unless the matrix is very large.
Note that the version of the task-based QR algorithm discussed in \tcite{D26} did not include most of the task insertion order and task priority optimizations that are discussed in Section \ref{sec:task_qr}.
Furthermore, as will be discussed in Subsection \ref{subsection:scalapack_conf}, the ScaLAPACK implementation used in \tcite{D26} did not allow the use of sufficiently large distributed blocks.

\subsection{Parallel bulge chasing}

\begin{figure}[h]
 \centering
 \includegraphics[scale=\figurescale]{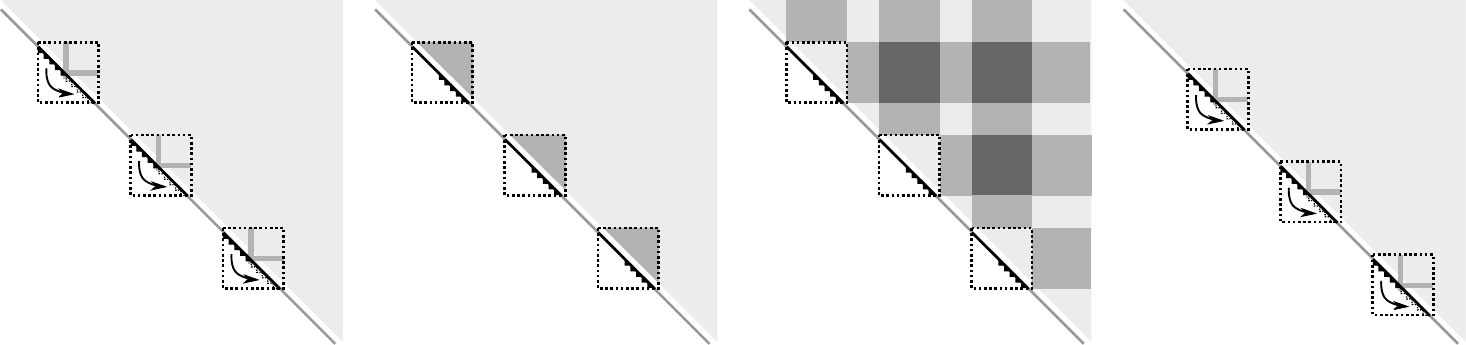}
 \caption{An illustration of multiple concurrent near-diagonal bulge chasing windows.
 This example includes three bulge chasing window chains.
 Note that the off-diagonal updates overlap each other.}
 \label{fig:multi_shift_scalapack}
\end{figure}

The set of bulges can be divided into several subsets and each subset can be associated with a corresponding bulge chasing window chain as shown in Fig. \ref{fig:multi_shift_scalapack}.
This can be used to (i) increase the concurrency, by the introduction of several concurrent diagonal windows as done in ScaLAPACK, and (ii) improve the arithmetical intensity, by the reduction of the size of each diagonal window so that it fits entirely inside the L1/L2 cache of a CPU core.

\subsubsection{ScaLAPACK implementation}

Note that the ScaLAPACK algorithm does not perform the near-diagonal bulge chasing and the off-diagonal updates as synchronously as implied in Fig. \ref{fig:multi_shift_scalapack}.
ScaLAPACK organizes the MPI processes into a two-dimensional process grid.
The MPI processes that are on the diagonal of the process grid perform the near-diagonal bulge chasing and then broadcast the accumulator matrices to the remaining MPI processes.
This is done using process row and process column broadcasts and therefore partially synchronizes the MPI processes.
Under optimal conditions, all MPI processes are involved when the off-diagonal updates are propagated.
The matrices are divided into rectangular blocks and distributed in a two-dimensional block cyclic fashion, see \tcite{blacs}.
The implementation synchronizes when the bulges are moved from one block to another.

\subsection{Aggressive early deflation}\label{subsec:aed}

\begin{figure}[h]
 \centering
 \includegraphics[scale=\figurescale]{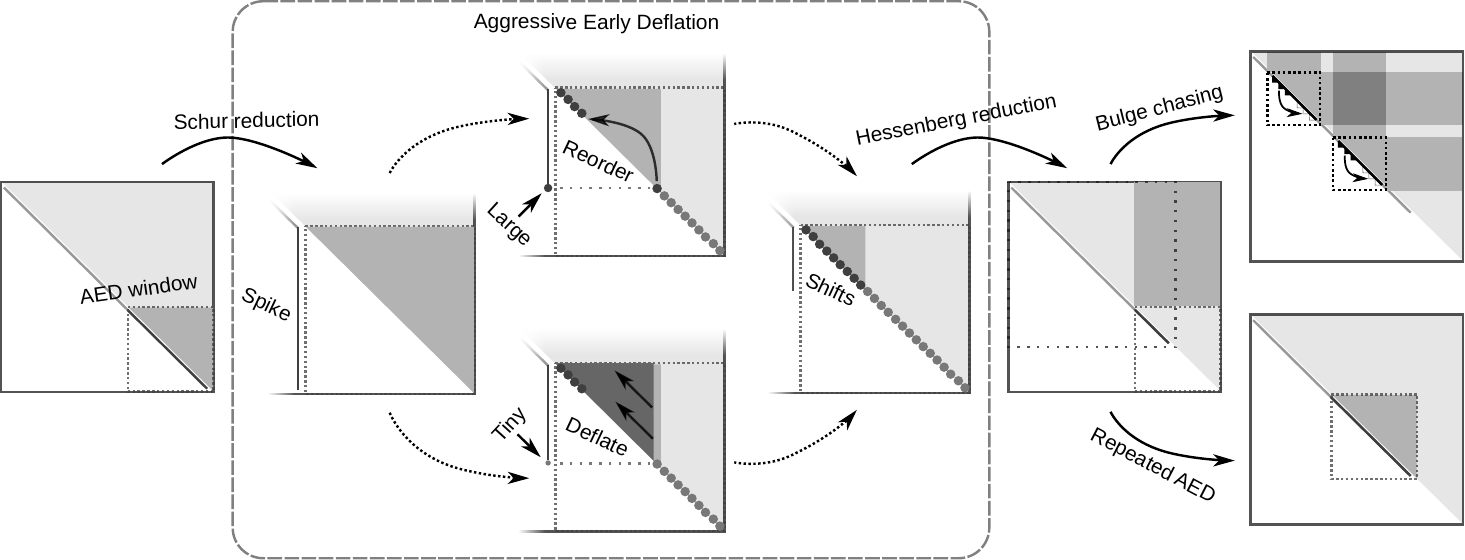}
 \caption{An illustration of the second iteration of the multi-shift QR algorithm with Aggressive Early Deflation (AED).}
 \label{fig:multi_shift_aed}
\end{figure}
 
The tightly-coupled multi-shifts are usually combined with the AED technique, see \tcite{Braman2002a}.
The purpose of an AED step is twofold: 
(i) attempt to detect deflatable eigenvalues early in order to accelerate the convergence rate of the QR algorithm and 
(ii) generate the shifts that are used to introduce the next set of tightly-coupled bulges.
Each AED step consists of three sub-steps as illustrated in Fig. \ref{fig:multi_shift_aed} and summarized below:
\begin{enumerate}
 \item A small diagonal window (a.k.a \emph{AED window}) in the lower-right corner of an unreduced block is reduced to (real) Schur form via recursive application of the QR algorithm.
 The related left-hand side updates induce a \emph{spike} to the left of the window and the upper Hessenberg structure is therefore temporarily lost.
 \item Computed \emph{eigenvalue candidates} (i.e., the eigenvalues of the involved AED window) are then systematically evaluated, starting from the bottommost diagonal block.
 If the corresponding element(s) in the same row(s) of the spike are found to be small enough (i.e., the so-called \emph{deflation condition} is satisfied), then the element(s) in the spike are set to zero and the eigenvalue or the complex conjugate pair of eigenvalues is thus deflated.
 If, on the other hand, the corresponding element(s) of the spike are found to be too large, then the AED window is reordered such that the diagonal block that failed the deflation check is moved above the remaining unevaluated diagonal blocks.
 This \emph{eigenvalue reordering} procedure pushes the remaining unevaluated diagonal blocks down the diagonal.
 The procedure involves an overlapping sequence of Givens rotations and 3-by-3 Householder reflectors; see \tcite{Bai1993a} and \tcite{Kagstrom1993}.
 \item After all eigenvalue candidates have been evaluated, the remaining spike is eliminated by performing a small-sized Hessenberg reduction.
\end{enumerate}
The failed eigenvalue candidates are used as shifts during the bulge chasing step, and if the AED step did not produce enough shifts, it is repeated.
The decision to repeat an AED step is also sometimes made for heuristic reasons since a large number of deflated eigenvalues usually implies that the next AED step is also likely to deflate a reasonable number of eigenvalues.
It should be noted that the AED window is copied to a separate memory buffer and copied back only when at least one eigenvalue was successfully deflated.
All orthogonal transformations that are initially applied only inside the AED window are later propagated as matrix-matrix multiplications.

\subsubsection{ScaLAPACK implementation}

The ScaLAPACK algorithm performs each AED step in parallel using a subset of the available MPI processes.
All MPI processes are effectively synchronized before and after each AED step.
The scalability of the improved ScaLAPACK implementation (see Subsection \ref{subsection:scalapack_conf}) is discussed and modeled in \citeauthor{GraKagKreShao2015a} \citeyearpar{GraKagKreShao2015a, GraKagKreShao2015b}. 
Also, as mentioned in Subsection \ref{subsec:multishifts}, the deliverable report \tcite{D26} contains a preliminary scalability comparison between LAPACK, ScaLAPACK and an early version of the task-based QR algorithm.

\section{Task-based QR algorithm}\label{sec:task_qr}

Design of a task-based algorithm begins with carefully dividing the computational work into self-contained tasks that all have well-defined inputs and outputs.
The way this differs from the common practice of organizing an implementation into regular subroutines is that a task-based algorithm relies on a \emph{runtime system}.
The task-based implementation does not call the associated computation kernels directly, instead it is the role of the runtime system to schedule the tasks to various computational resources, such as CPU cores and GPUs, in a \emph{sequentially consistent order} as dictated by the task dependencies that arise from the inputs and the outputs.
One of the main benefits of this approach is that as long as the task-based algorithm is well-designed, the underlying parallelism is exposed automatically as the runtime system gradually traverses the resulting \emph{task graph}.

In particular, the fact that the runtime system guarantees that the tasks are executed in a sequentially consistent order eliminates the need to explicitly synchronize the execution. 
Different computational steps, that were previously separated by global synchronization points, are allowed to merge and the runtime system readily exploits the available concurrency to a far higher degree than previously.
Other benefits of the task-based approach include, for example, better load balancing and resource utilization due to dynamic task scheduling.
This all leads to significantly more powerful algorithms and implementations that are able to dynamically adapt to different input data and ever-changing hardware configurations.

\subsection{StarPU runtime system}\label{subsec:starpu}

The implementation of the task-based QR algorithm is built on top of the StarPU runtime system, see \tcite{starpu}.
The following subsections provide a brief introduction to StarPU.

\subsubsection{Tasks and codelets}

A task-based implementation defines a set of \emph{codelets} that specify how each \emph{task type} is implemented.
Each task type can have multiple implementations defined in the corresponding codelet, for example, one implementation for a CPU core and another implementation for a GPU.
StarPU uses calibrated \emph{performance models} to predict the execution and data transfer times for each computational resource, and makes scheduling decisions based on that information.
Each task can also be given an integer \emph{priority} that affects the scheduling after all task's dependencies have been executed.
Larger number implies higher priority.

\subsubsection{Data handles and task dependencies}

The input and output data is encapsulated inside \emph{data handles}.
In particular, the StarPU-based implementation divides the matrices into disjoint \emph{tiles} that are square with the possible exception of the last tile row and tile column. 
Each tile is registered with StarPU and given a data handle.
In our chosen configuration, StarPU derives the task dependencies from the input and output data handles.
That is, the task graph is constructed implicitly.
An application controlled \emph{main thread} inserts the tasks into StarPU and StarPU schedules the tasks to a set of worker threads.
If the main thread needs the output of a task, it must \emph{acquire} the corresponding data handle.
This causes the main thread to wait until the task has been executed.

\subsubsection{Multi-node environments}

StarPU supports multi-node environments (distributed memory) though MPI.
In our chosen configuration, each data handle is supplemented with an unique \emph{tag} (integer identifier) and tasks that require communication are inserted in all involved MPI processes.
In addition, each data handle has an \emph{owner}, i.e., the MPI process where the master copy of the data is located.
Other MPI processes register a placeholder data handle that may encapsulate a copy of the data during the computations.
The task graph, the tags, and the owner information provide StarPU all necessary information for automatically performing the communications.
In a sense, each MPI process has a (partial) copy of the task graph that has been supplemented with special communication tasks.
A separate worker thread is dedicated to executing the communication tasks using MPI.
Note that the master thread must insert \emph{explicit communication requests} into StarPU in some situations, for example, when it is about to acquire a data handle.

\subsubsection{Distributed blocks}\label{subsection:distr_blocks}

In a multi-node configuration, the StarPU-based implementation adds an additional layer of structure on top of the tiling.
The tiles are consolidated into disjoint rectangular \emph{blocks} of uniform size (excluding the last block row and block column).
All tiles that belong to the same \emph{distributed block} must have the same owner.
Beyond that, the data distribution can be arbitrary.

\subsection{Task types}

The task-based algorithm is built around seven main task types:

\begin{description}
 
 \item[{\tt small Schur}] task reduces a (small) unreduced block to Schur form.
 The local transformations are accumulated into an accumulator matrix and the task returns the outcome of the reduction operation (either success or failure).
 The task is implemented as a call to the \texttt{DHSEQR} LAPACK routine (sequential multi-shift QR algorithm with AED).
 
 \item[{\tt small AED}] task performs a sequential AED inside a diagonal window.
 The local transformations are accumulated into an accumulator matrix and the task returns the outcome of the AED (the number of deflated eigenvalues) and the computed shifts.
 The Hessenberg reduction sub-step is part of the {\tt small AED} task.
 Future work includes the separation of this sub-step into a separate task.
 The task implementation uses the \texttt{DHSEQR}, \texttt{DTREXC} (eigenvalue reordering), \texttt{DGEHRD} (Hessenberg reduction) and \texttt{DORMHR} (Hessenberg reduction) LAPACK routines.
 
 \item[{\tt push bulges}] task chases a set of tightly-coupled bulges down the diagonal inside a diagonal window.
 Optionally, the task introduces and/or annihilates the bulges. 
 The local transformations are accumulated into an accumulator matrix.
 Those {\tt push bulges} tasks, that are associated with the last set of tightly-coupled bulges, finish by scanning the sub-diagonal entries in order to identify any unreduced blocks that were decoupled during the bulge chasing step. 
 This decoupling can happen either due to vigilant deflations or entries being flushed to zero.
 The outcome of this sub-diagonal scan is stored to a special \emph{aftermath vector} that has an entry for each row of the matrix.
 The task implementation chases the bulges in small sub-batches and utilizes reflector accumulation and BLAS-3 operations.   
 Each diagonal window is placed such that its lower-right corner follows the edges of the underlying tiles.
The size of each bulge chasing window is at most $2b_{tile} \times 2 b_{tile}$, where $b_{tile}$ is the tile size, and each task chases at most $(b_{tile}-1)/3$ bulges.
 {\tt push bulges} tasks that are associated with the same set of tightly-coupled bulges form a bulge chasing window chain.
 An earlier version of the implementation (see \tcite{D26}) reused some sequential ScaLAPACK code, but this code has since been replaced with code that has been written from the scratch in C.
 
 \item[{\tt deflate}] task attempts to deflate eigenvalue candidates that fall within a diagonal window and reorders failed candidates to the upper-left corner of the window.
 {\tt deflate} tasks are also used to reorder a set of failed eigenvalue candidates to the upper-left corner of the AED window.
 In the reordering configuration, {\tt deflate} tasks that are associated with the same set of failed eigenvalue candidates form a \emph{reordering window chain}.
 The local transformations are accumulated into an accumulator matrix and the task returns the outcome of the deflation check (the locations of the topmost and bottommost failed eigenvalue candidates).
 Each diagonal window is placed such that its upper-left corner follows the edges of the underlying tiles.
The size of each deflation window is at most $2b_{tile} \times 2 b_{tile}$ and each task moves at most $b_{tile}-1$ eigenvalues.
 The task implementation uses the \texttt{DTREXC} LAPACK routine.
 
 \item[{\tt small Hessenberg}] task reduces a diagonal window to a Hessenberg form.
 The local transformations are accumulated into an accumulator matrix.
 The task is implemented as calls to the \texttt{DGEHRD} and \texttt{DORMHR} LAPACK routines.
 
 \item[{\tt left update}] task applies an accumulator matrix from the left to a section of a matrix. 
 See Subsection \ref{subsec:update_tasks}.
 
 \item[{\tt right update}] task applies an accumulator matrix from the right to a section of a matrix.
 See Subsection \ref{subsec:update_tasks}.
 
\end{description}

The task-based algorithm contains several auxiliary task types and the StarNEig library includes a task-based implementation of the Hessenberg reduction phase which is used to eliminate spikes that are larger than a given threshold.
For most task types, the implementation copies the contents of the intersecting tiles into a temporary buffer before the computations. 
The return values are stored to separate data handles and later acquired by the main thread.

\begin{remark}
 In the generalized case, the task-based algorithm includes one additional task type:
 \begin{description}
  \item[{\tt push infinities}] task chases a set of infinite eigenvalues up the diagonal inside a diagonal window.
  The local transformations are accumulated into an accumulator matrix.
  The chasing is performed with a sequence of Givens rotations (see, e.g., \tcite{Adlerborn2014}).
  Each diagonal window is placed such that its upper-left corner follows the edges of the underlying tiles.
The size of each diagonal window is at most $2b_{tile} \times 2 b_{tile}$ and each task chases at most $b_{tile}-1$ infinite eigenvalues.
 {\tt push infinities} tasks that are associated with the same set of infinite eigenvalues form an infinite eigenvalue chasing window chain.
 \end{description}
 
The other task types differ as follows:
\begin{description}
 
 \item[{\tt small Schur}] task implements a sequential version of multi-shift QZ algorithm with AED.
 The implementation uses the \texttt{DHGEQZ} (double-implicit-shift QZ algorithm), \texttt{DTGEXC} (generalized eigenvalue reordering) and \texttt{DGGHRD} (Hessenberg-triangular reduction) LAPACK routines.
 
 \item[{\tt small AED}] task implementation uses the same sequential QZ algorithm as the {\tt small Schur} task.
 
 \item[{\tt push bulges}] task attempts to detect any infinite eigenvalues.
 Discovered infinite eigenvalues are marked in the aftermath vector.
 
 \item[{\tt deflate}] task implementation uses the \texttt{DTGEXC} LAPACK routine.
 
 \item[{\tt small Hessenberg}] task is implemented as a call to the \texttt{DGGHRD} LAPACK routine.
 
\end{description}
\end{remark}

\subsubsection{Update tasks}\label{subsec:update_tasks}

\begin{figure}[h]
 \centering
 \includegraphics[scale=\figurescale]{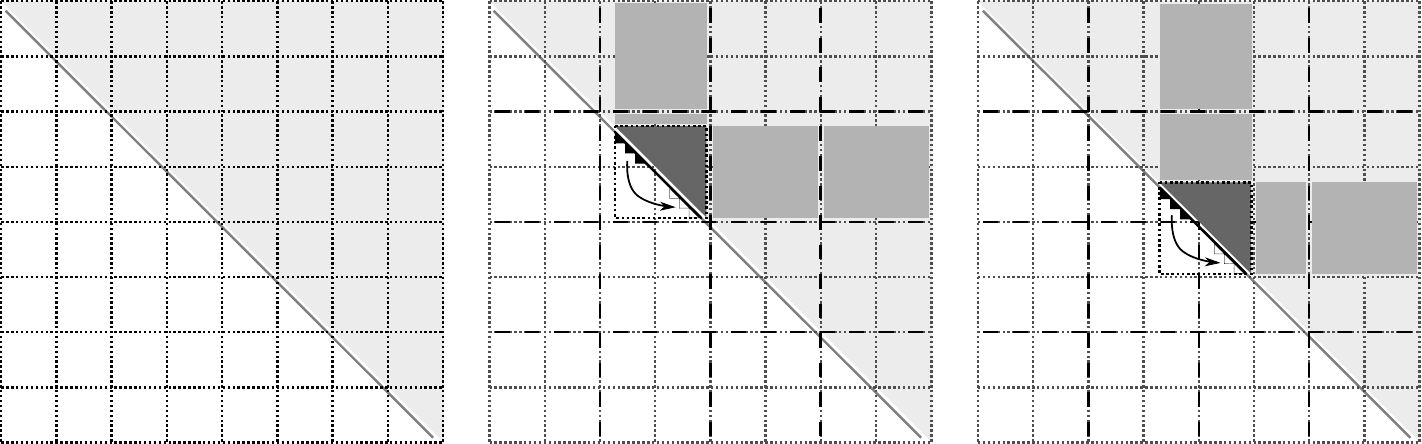}
 \caption{
 An illustration of how a matrix is divided into square tiles and how the left-hand and right-hand side updates are cut into individual update tasks using a 4-by-4 stencil.
 }
 \label{fig:splicing}
\end{figure}

We will from now on refer to the {\tt left update} and {\tt right update} tasks as \emph{update tasks}.
The update tasks are implemented as a call to the \texttt{dgemm} BLAS routine (for CPUs) or the \texttt{cublasDgemm} cuBLAS routine (for GPUs).  
Special care has to be taken when cutting the updates into individual update tasks or we risk introducing spurious data dependencies between independent tasks.
For example, if two {\tt left update} tasks, that were cut from the same left-hand side update, were allowed to operate on a common tile, then this overlap would induce a spurious dependency between the two tasks. 
The task-based algorithm avoids this problem by forming a two-dimensional stencil (see Fig. \ref{fig:splicing}) that separates the tiles into groups of adjacent tiles and the updates are always cut following the edges of the stencil.
In a multi-node environment, the stencil is formed such that it follows the edges of the underlying distributed blocks.
In this way, the implementation avoids spanning an update task across multiple MPI processes.

\subsection{Event-driven algorithm}\label{subsec:eventdriven}

The QR algorithm introduces several challenges for the task-based approach, the most significant of them being the fact that the QR algorithm is iterative and contains several data-dependent branching points.
For example, it is practically impossible to predict how many eigenvalues each AED step manages to deflate and this information is critical when deciding whether to perform a repeated AED step or to proceed to a bulge chasing step.
This means that the task-based algorithm cannot simply insert all tasks to the runtime system and then await their completion.
This is a major difference to, say, Cholesky factorization (see, e.g., \tcite{10.1007/s10766-016-0441-6}) or eigenvalue reordering (see \tcite{Myllykoski2018} and \tcite{NLAFET-WN11}).
Instead, the task-based algorithm must observe the outcomes of certain tasks in order to make the necessary decisions and this can lead to an early exhaustion of the task pool since no new tasks can be inserted while the main thread is being blocked inside the related API function of the runtime system.

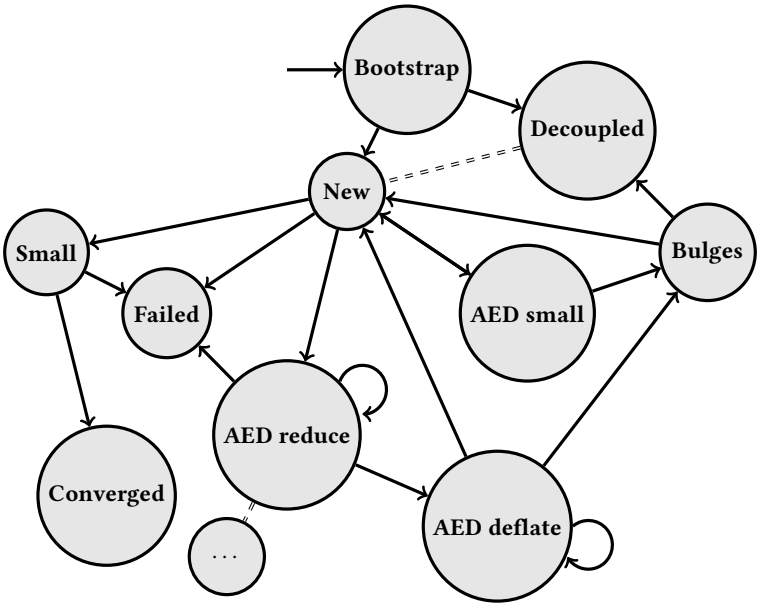
\begin{figure}[h]
  \centering
  \small
  \begin{tikzpicture}
    [scale=0.8,
    y=-1cm,
    arrow/.style={very thick,->},
    state/.style={fill=gray!20,draw,very thick,circle,minimum width=10mm}]
    
    \node [state] (bootstrap) at (1,0) {\bf Bootstrap};
    \node [state] (children) at (4,1) {\bf Decoupled};
    \node [state] (new) at (0,2) {\bf \bf New};
    \node [state] (failed) at (-3,4) {\bf Failed};
    \node [state] (small) at (-5,3) {\bf Small};
    \node [state] (aed_small) at (3.0,4.0) {\bf AED small};
    \node [state] (aed_schur) at (-1,6) {\bf AED reduce};
    \node [state] (aed_deflate) at (2.5,7.5) {\bf AED deflate};
    \node [state] (bulges) at (6,3) {\bf Bulges};
    \node [state] (converged) at (-4,7) {\bf Converged};
    
    \coordinate (START) at (-1,0);
    \draw [arrow] (START) to (bootstrap);
    
    \node [state] (ITER) at (-2,8) {$\dots$};
    \draw [double,dashed] (aed_schur) to (ITER);
    
    \draw [arrow] (bootstrap) to (new);
    \draw [arrow] (bootstrap) to (children);
    
    \draw [arrow] (new) to (small);
    \draw [arrow] (new) to (aed_small);
    \draw [arrow] (new) to (aed_schur);
    \draw [arrow] (new) to (failed);
    
    \draw [arrow] (small) to (failed);
    \draw [arrow] (small) to (converged);
    
    \draw [arrow] (aed_small) to (new);
    \draw [arrow] (aed_small) to (bulges);
    
    \draw [arrow] (aed_schur) to (aed_deflate);
    \draw [arrow] (aed_schur) to (failed);
    
    \draw [arrow] (aed_deflate) to (new);
    \draw [arrow] (aed_deflate) to (bulges);
    
    \draw [arrow] (bulges) to (new);
    \draw [arrow] (bulges) to (children);
    
    \draw [double,dashed] (children) to (new);
    
    \draw [arrow] (aed_schur.north east)arc(-160:90:0.4);
    \draw [arrow] (aed_deflate.east)arc(-130:150:0.4);

  \end{tikzpicture}
  
  \caption{Possible segment states and state transitions.
  The transitions to processing a child-segment list are illustrated with dashed double lines.}
  \label{fig:states}
\end{figure}

Our solution to these problems is to adopt an \emph{event driven approach} and eliminate as many decision points as possible.
All information on the unreduced blocks is stored in a nested list where each element corresponds to one unreduced block.
We will from now on refer to the list elements as \emph{segments}.
Each segment has a \emph{state} and can contain a \emph{child-segment list}.
Each segment list also contains a set of preset \emph{priority levels} ({\tt max\_prio}, {\tt def\_prio} and {\tt min\_prio}) that are used to assign priorities for tasks that are related to the segment list.
The task-based algorithm repeatedly scans the segment list and performs an \emph{action} on each segment as illustrated in Fig. \ref{fig:states} and defined below:
\begin{description}

 \item[Bootstrap] state initializes the algorithm.
 \begin{description}
  \item[\it Initialization:] 
  A new segment covering the entire matrix $H$ is created, initialized with state {\bf Bootstrap} and added to the segment list. 
  The sub-diagonal entries of $H$ are scanned in order to detect any preexisting unreduced blocks and this information is communicated to all MPI processes in the form of an aftermath vector.
  The segment list priorities are set as follows:
  \begin{itemize}
   \item The highest task priority ({\tt max\_prio}) is set to the maximum priority allowed by the runtime system.
   \item The default task priority ({\tt def\_prio}) is set to the runtime system's default priority. 
   \item The lowest task priority ({\tt min\_prio}) is set to the minimum priority allowed by the runtime system.
  \end{itemize}
  \item[\it Action:] 
  Each MPI process piecewise acquires and processes the aftermath vector.
  (i) If a new unreduced block is detected, then a new child-segment corresponding to the new unreduced block is created, initialized with state {\bf New} and added to the child-segment list.
  At this point, an action is also triggered for the new child-segment.
  The parent segment is marked as {\bf Decoupled}.
  Note that several child-segments can be created during the scan.
  (ii) Otherwise, the segment is marked as {\bf New}.
 \end{description}
 
 \item[New] state indicates the beginning of a new iteration.
 \begin{description}
  \item[\it Action:]
  (i) If the iteration limit is reached, then the segment is marked as {\bf Failed}. 
  (ii) Otherwise, if the segment is small, then a {\tt small Schur} and related update tasks are inserted, and the segment is marked as {\bf Small}. 
  (iii) Otherwise, if the AED window is small, then a {\tt small AED} task is inserted and the segment is marked as {\bf AED small}.
  (iv) Otherwise, the AED window is copied to a separate matrix and a new child-segment list is created for it. 
  A new child-segment covering the entire copied matrix is created, initialized with state {\bf New}, and added to the child-segment list. 
  The parent segment is marked as {\bf AED reduce}.
  The child-segment list priorities are set as follows:
  \begin{itemize}
   \item The highest task priority ({\tt max\_prio}) is set to {\tt max\_prio\textsubscript{parent}}.
   \item The lowest task priority ({\tt min\_prio}) is set to {\tt min(max\_prio, def\_prio\textsubscript{parent} + 1)}.
   \item The default task priority ({\tt def\_prio}) is set to {$\lfloor$(\tt max\_prio + min\_prio) / 2$\rfloor$}. 
  \end{itemize}
  See Subsection \ref{subsec:parallel_aed} for further information relating to the AED step.
 \end{description}
 
 \item[Small] state indicates that a {\tt small Schur} task has been inserted.
 \begin{description}
  \item[\it Action:] 
  The return status of the {\tt small Schur} task is acquired.
  (i) If the {\tt small Schur} task was executed successfully, then the segment is marked as {\bf Converged}. 
  (ii) Otherwise, the segment is marked as {\bf Failed}.
 \end{description}
 
 \item[Failed] state indicates that the algorithm failed to reduce the segment to Schur form.
 \begin{description}
  \item[\it Action:] The task-based algorithm exits and returns an appropriate error code.
 \end{description}
 
 \item[Converged] state indicates that the segment was successfully reduced to Schur form.
 \begin{description}
  \item[\it Action:] The segment is removed from the segment list.
 \end{description}
 
 \item[Decoupled] state indicates that the segment has decoupled into several child-segments.
 \begin{description}
  \item[\it Action:] The parent segment is replaced with the child-segments.
 \end{description}
 
 \item[AED small] state indicates that a {\tt small AED} task has been inserted.
 \begin{description}
  \item[\it Action:] 
  The return status of the {\tt small AED} task is acquired.
  If the {\tt small AED} task deflated one or more eigenvalues, then the AED window is embedded back into the matrix and the related update tasks are inserted. 
  (i) If the {\tt small AED} task generated enough shifts, then the bulge chasing tasks are inserted (see Subsection \ref{subsec:bulges}) and the segment is marked as {\bf Bulges}.
  Communication requests for communicating the aftermath vector to all MPI processes are inserted.
  (ii) Otherwise, the segment is marked as {\bf New}.
 \end{description}
 
 \item[AED reduce] state indicates that the associated AED window is being reduced to Schur form.
 \begin{description}
  \item[\it Action:] 
  An action is triggered for each segment in the child-segment list.
  (i) If any of the child-segments transitions to the {\bf Failed} state, then the parent segment is marked as {\bf Failed}.
  (ii) Otherwise, if the child-segment list is non-empty, then the parent segment retains the {\bf AED reduce} state.
  (iii) Otherwise, the first {\tt deflate} task and the related update tasks are inserted, and the parent segment is marked as {\bf AED deflate}.
  See Subsection \ref{subsec:parallel_aed}.
 \end{description}
 
 \item[AED deflate] state indicates that the associated AED window is being deflated.
 \begin{description}
  \item[\it Action:] 
  The return status of the {\tt deflate} task is acquired.
  (i) If more {\tt deflate} tasks are required, then the next {\tt AED deflate} task and the related update tasks are inserted, and the segment retains the {\bf AED deflate} state. 
  If necessary, the reordering related {\tt deflate} tasks and the related update tasks are also inserted.
  (ii) Otherwise, if enough shifts have been generated, then the bulge chasing tasks are inserted (see Subsection \ref{subsec:bulges}) and the segment is marked as {\bf Bulges}.
  Communication requests for communicating the aftermath vector to all MPI processes are inserted.
  (iii) Otherwise, the segment is marked as {\bf New}.
  In both cases (ii) and (iii), if one or more eigenvalues have been deflated, then the AED window is first embedded back into the matrix and the related update tasks are inserted.
  See Subsection \ref{subsec:parallel_aed} for further information relating to the AED step.
 \end{description}
 
 \item[Bulges] state indicates that the bulge chasing tasks have been inserted.
 \begin{description}
  \item[\it Action:] 
  Each MPI process piecewise acquires and processes the aftermath vector.
  (i) If a new unreduced block is detected, then a new child-segment corresponding to the new block is created, initialized with state {\bf New} and added to a child-segment list.
  At this point, an action is also triggered for the new child-segment.
  The parent segment is marked as {\bf Decoupled}.
  Note that several child-segments can be created during the scan.
  (ii) Otherwise, the segment is marked as {\bf New}.
 \end{description}
 
\end{description}

\begin{figure}[h]
 \centering
 \includegraphics[scale=\figurescale]{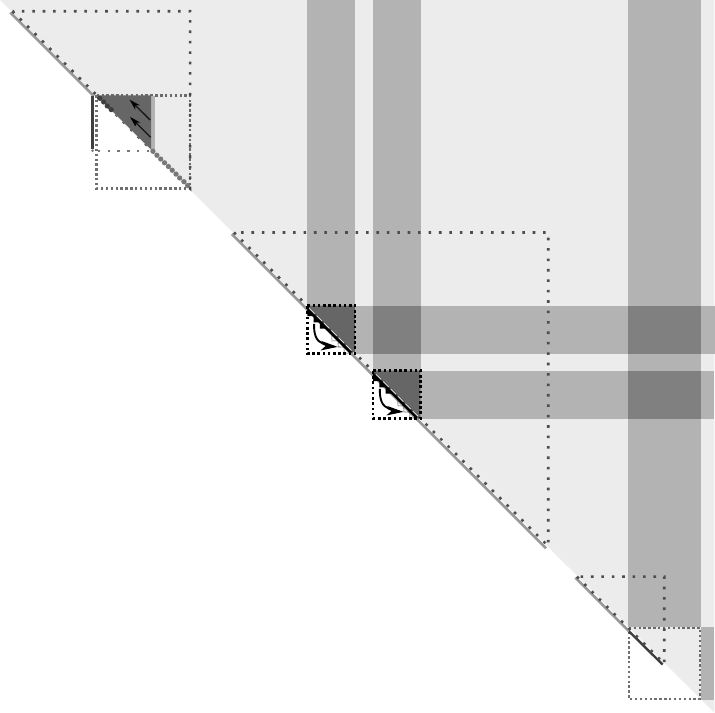}
 \caption{
 An illustration of three unreduced blocks (segments) that all are in different states.
 From top to bottom: 
 (i) 2nd AED sub-step ({\bf AED small} or {\bf AED deflate}), 
 (ii) bulge chasing ({\bf Bulges}), and 
 (iii) off-diagonal updates from an AED step ({\bf Bulges}, {\bf AED small} or {\bf AED reduce}).
 }
 \label{fig:segments}
\end{figure}

As illustrated in Fig. \ref{fig:segments}, the task-based algorithm allows several segments to be processed concurrently.
In particular, the task-based algorithm can detect new segments even when the bulge chasing step is still active.
The main thread still waits for the completion of certain tasks because the StarPU-based implementation uses the blocking variant of the data handle acquiring function.
The non-blocking variant was tested but the additional overhead (the main thread constantly pooling the runtime system) and, more importantly, complications relating to the ordering of operations in multi-node configurations lead us to abandon this approach.

\begin{remark}
 In the generalized case, the {\bf Bootstrap} and {\bf Bulges} actions also scan the aftermath vector for infinite eigenvalues. 
 If infinite eigenvalues are discovered, they are grouped, chased up the diagonal and deflated. 
 A similar approach was used in \tcite{Adlerborn2014}.
 Once one group fills (at most $b_{tile}-1$ infinite eigenvalues), the related {\tt push infinities} and update tasks are inserted.
 The task insertion is also triggered if the segment decouples into sub-segments.
 This approach makes it possible to overlap an infinite eigenvalue chasing step with a bulge chasing step.
 The infinite eigenvalues are chased to the upper-left corner of the unreduced block because in this way the involved tasks do not block the next AED step.
\end{remark}

\subsection{Bulge chasing}\label{subsec:bulges}

Although the vast majority of the flops associated with a bulge chasing step are performed using BLAS-3 operations, the overall cost of a bulge chasing step is still very high in comparison to an AED step.
In addition, each bulge chasing step modifies a significant fraction of the entire matrix.
It is therefore important to optimize the bulge chasing step as much as possible.
In particular, as demonstrated in \tcite{Myllykoski2018} and \tcite{NLAFET-WN11}, the order in which the tasks are inserted can play a major role in the performance of two-sided matrix transformation algorithms.
The order in which the tasks are inserted effectively determines the order in which the orthogonal transformations are applied.

\subsubsection{Critical path}

In this section, we use $\mathcal{G}_{qr}$ to denote the subgraph that forms the \emph{critical path} of the entire QR algorithm.
By a critical path, we mean the longest path through the tasks graph when measured in terms of the execution time.
The critical path therefore gives a lower bound for the execution time of the algorithm.
We say that a task \emph{feeds back to} the critical path when the task is a direct or indirect dependency to one or more tasks on the critical path.
We also use $\mathcal{G}_{bulges}$ to denote the subgraph that forms the critical path of the bulge chasing step that is currently under consideration.
The subgraphs $\mathcal{G}_{qr}$ and $\mathcal{G}_{bulges}$ are not known.
However, we can postulate the following: 
(i) All {\tt push bulges} tasks must be executed before the next AED step can begin.
All {\tt push bulges} tasks are therefore part of both $\mathcal{G}_{qr}$ and $\mathcal{G}_{bulges}$.
(ii) All AED-related tasks that contribute to the computation of the shifts must be executed before the next bulge chasing step can begin.
Many AED-related tasks are therefore part of $\mathcal{G}_{qr}$.
(iii) The subgraph $\mathcal{G}_{bulges}$ is a subgraph of $\mathcal{G}_{qr}$.

\begin{figure}[h]
 \centering
 \includegraphics[scale=\figurescale]{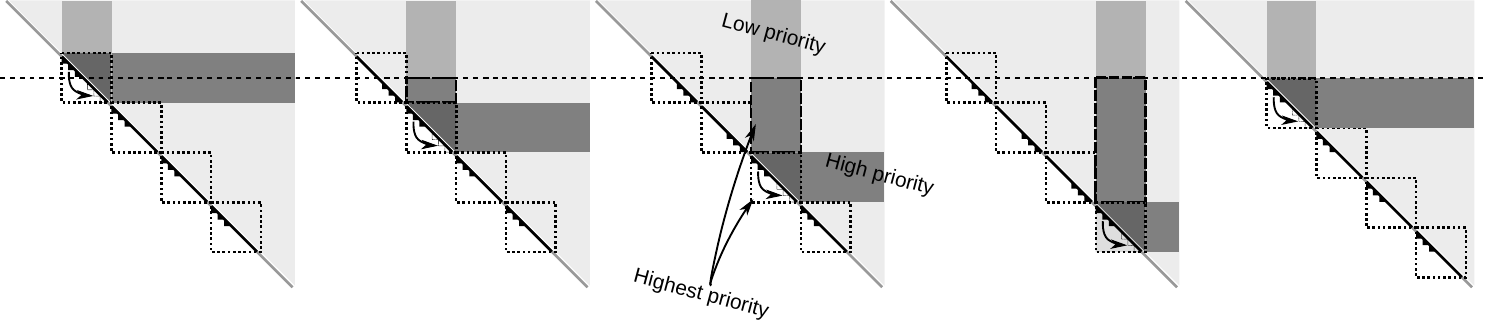}
 \caption{
 An illustration of a task insertion step.
 The task insertion order is read from left to right.
 All right-hand side updates above the dashed line are considered low priority as the associated {\tt right update} tasks do not feed back to $\mathcal{G}_{bulges}$.
 This example involves four bulge chasing window chains.}
 \label{fig:windows}
\end{figure}

\subsubsection{Task insertion step}

When the tasks are inserted, all bulge chasing windows chains are processed together, in an interleaved manner, as visualized in Fig. \ref{fig:windows}.
During each \emph{task insertion step}, the diagonal bulge chasing windows are processed in reverse order\footnote{The most straightforward task insertion order would be to insert each bulge chasing window chain separately, starting from the first bulge chasing window chain. Compared to this straightforward task insertion order, the task-based algorithm inserts the bulge chasing window chains in an interleaved manner, in reverse order.} starting from the topmost window.
Both the {\tt push bulges} task and the corresponding {\tt left update} and {\tt right update} tasks are inserted.
The {\tt push bulges} task is always given the highest priority ({\tt max\_prio}) and the {\tt left update} tasks are given the second highest priority ({\tt max(def\_prio, max\_prio - 1)}).
The {\tt right update} tasks below the \emph{dashed line} (see Fig. \ref{fig:windows}) are given the highest priority ({\tt max\_prio}) and the {\tt right update} tasks above the dashed line are given the default priority ({\tt def\_prio}).
Those {\tt right update} tasks that update the $Q$ matrix are given the lowest priority ({\tt min\_prio}).

\subsubsection{Ordering and prioritization}

The overall goal is to always prioritize the {\tt push bulges} and {\tt left update} tasks since {\tt push bulges} tasks are part of $\mathcal{G}_{bulges}$ and {\tt left update} tasks feed back to $\mathcal{G}_{bulges}$.
In particular, since the bulge chasing window chains are processed in reverse order, the {\tt left update} tasks are always scheduled before the overlapping {\tt right update} tasks (see Fig. \ref{fig:windows}).
We could, in principle, insert the tasks in an arbitrary order, in two phases: first inserting the {\tt push bulges} and {\tt left update} tasks, and then the {\tt right update} tasks.
However, the simplest approach is to process the bulge chasing windows in the described manner.

\subsubsection{Bulge chasing windows chains and task graphs}

\begin{figure}[h]
 \centering
 \includegraphics[scale=\figurescale]{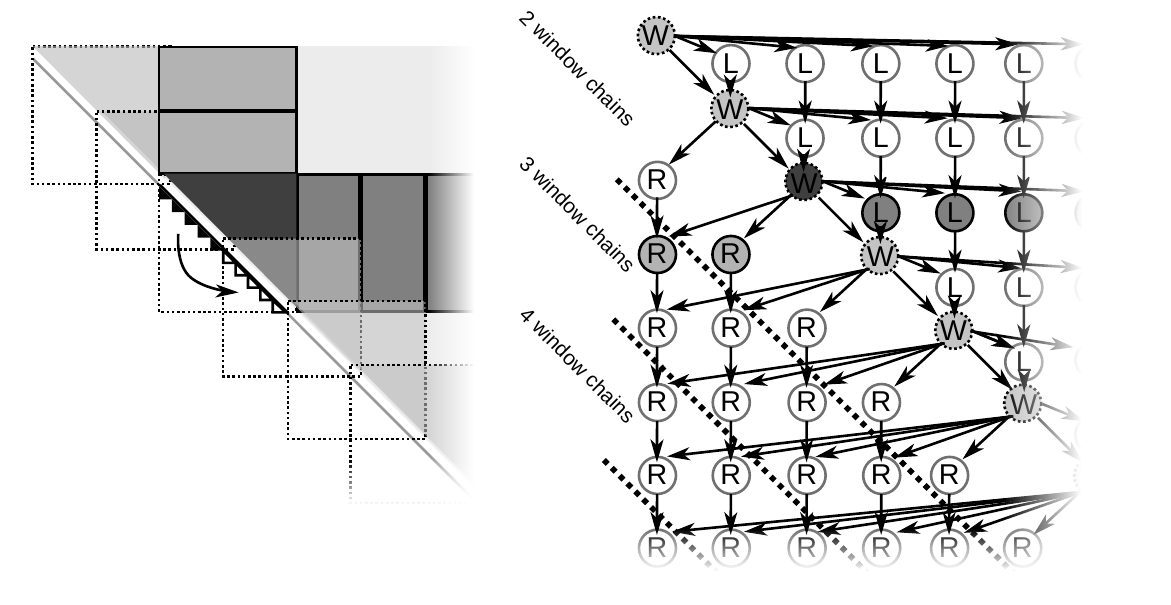}
 \caption{A simplified tasks graph for the first bulge chasing window chain.
 The {\tt push bulges} tasks are marked with W, {\tt left update} tasks with L and {\tt right update} tasks with R.
 The dashed lines show which tasks feed back to $\mathcal{G}_{bulges}$ for a given number of bulge chasing window chains.
 For example, ''4 window chains`` corresponds to the situation shown in Fig. \ref{fig:windows}, i.e., all tasks above the third dashed line feed back to $\mathcal{G}_{bulges}$.  
 }
 \label{fig:chain_flow}
\end{figure}

Note that all {\tt push bulges} and {\tt left update} tasks, that are inserted during the next task insertion step, are below the dashed line in Fig. \ref{fig:windows}.
This means that the {\tt right update} tasks, that update sections of the matrix above the dashed line, do not feed back to $\mathcal{G}_{bulges}$ and are thus given a lower priority.
The {\tt right update} tasks, that update sections of the matrix below the dashed line in Fig. \ref{fig:windows}, do feed back to $\mathcal{G}_{bulges}$ and are therefore given the highest priority.
This encourages the runtime system to schedule these tasks as soon as possible so that they do not block the tasks that are inserted during the next task insertion step.
Fig. \ref{fig:chain_flow} shows a simplified task graph for the first bulge chasing window chain.
The dashed lines indicate which tasks feed back to $\mathcal{G}_{bulges}$ for a given number of bulge chasing window chains.
For example, ''4 window chains`` corresponds to the situation shown in Fig. \ref{fig:windows}, i.e., all tasks above the third dashed line feed back to $\mathcal{G}_{bulges}$.

\begin{figure}[h]
 \centering
 \includegraphics[scale=\figurescale]{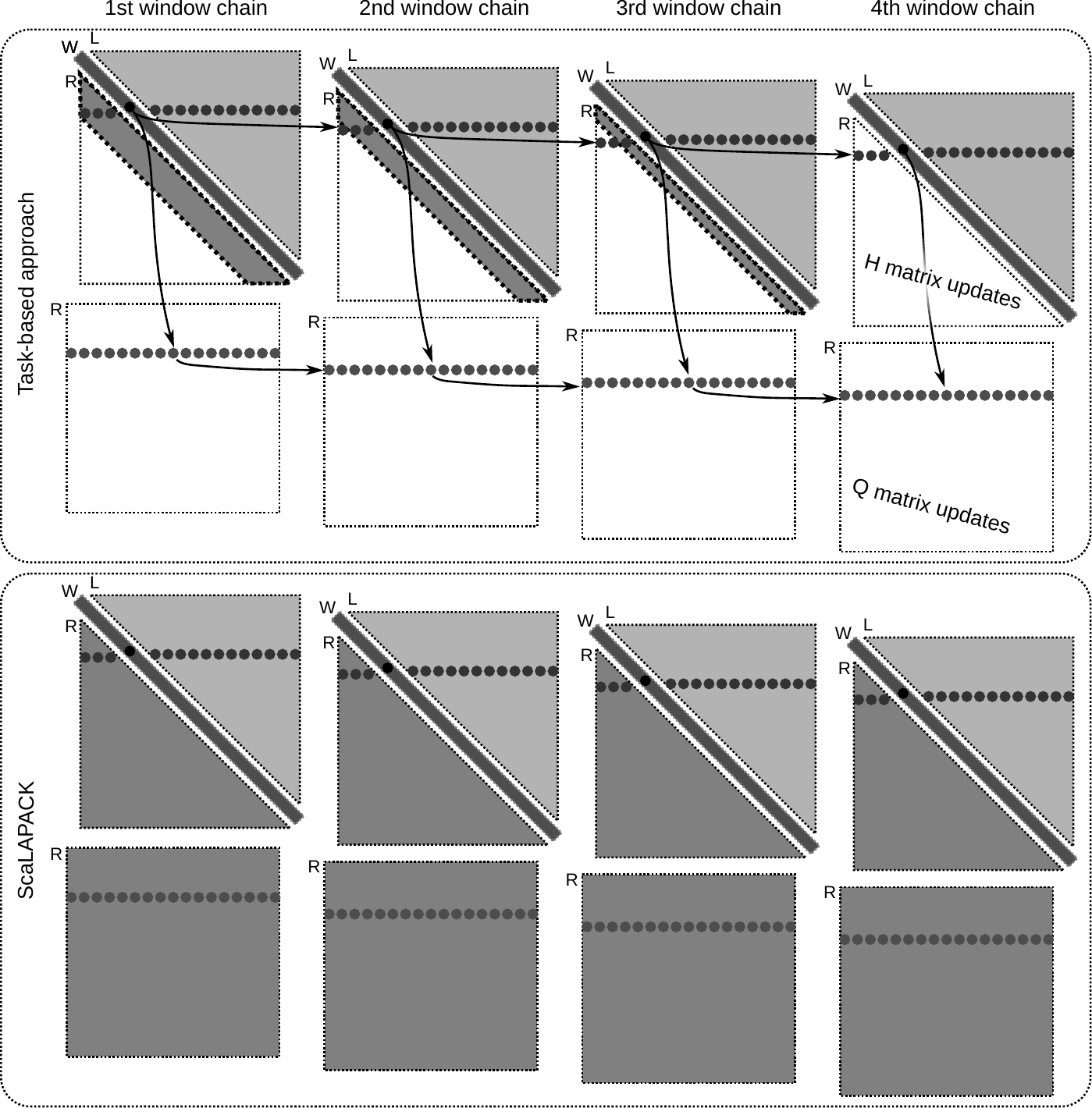}
 \caption{
 A simplified task graph that contains four bulge chasing window chains (compare to the right half of Fig. \ref{fig:chain_flow}).
 The sections of the task graph that need to be completed before the next AED step can begin are highlighted in various shades of gray.
 In particular, the {\tt right update} tasks that feed back to $\mathcal{G}_{bulges}$ are highlighted with \emph{dashed perimeters}.
 It is critical to realize that the right-hand side updates below the dashed line in Fig. \ref{fig:windows} correspond to the {\tt right update} tasks above the third dashed line in Fig. \ref{fig:chain_flow}.
 The same {\tt right update} tasks are highlighted with the dashed perimeters in this figure.
 If we had added more separation between the bulge chasing windows during the task insertion step, then sections highlighted with the dashed perimeters would be larger.
 For illustrational purposes, some circles correspond to two update tasks.
 }
 \label{fig:chain_flow2}
\end{figure}

\subsubsection{Shortening of the critical path}\label{subsec:shortening_path}

If we had added more separation between the bulge chasing windows during the task insertion step, then more {\tt right update} tasks would feed back to $\mathcal{G}_{bulges}$ as the distance from each bulge chasing window to the dashed line in Fig. \ref{fig:windows} would be longer.
Therefore, given the choice between zero or some separation between the windows, the most optimal configuration is the one where the diagonal windows are packed as close to each other as possible.
Note that real-world task graphs are significantly larger than the simplified task graph presented in Fig. \ref{fig:chain_flow}.
Therefore, as shown in Fig. \ref{fig:chain_flow2}, the {\tt right update} tasks that do not feed back to $\mathcal{G}_{bulges}$ form nearly half of the task graph.
Furthermore, {\tt right update} tasks that compute the $Q$ matrix form a completely separate subgraph and constitute about one-half of the total number of update tasks.
Therefore, as illustrated in Fig. \ref{fig:chain_flow2}, the number of update tasks that need to be executed before the next AED step can begin is reduced by almost three-quarters when compared to the ScaLAPACK algorithm. 

\begin{figure}[h]
 \centering
 \includegraphics[width=\textwidth]{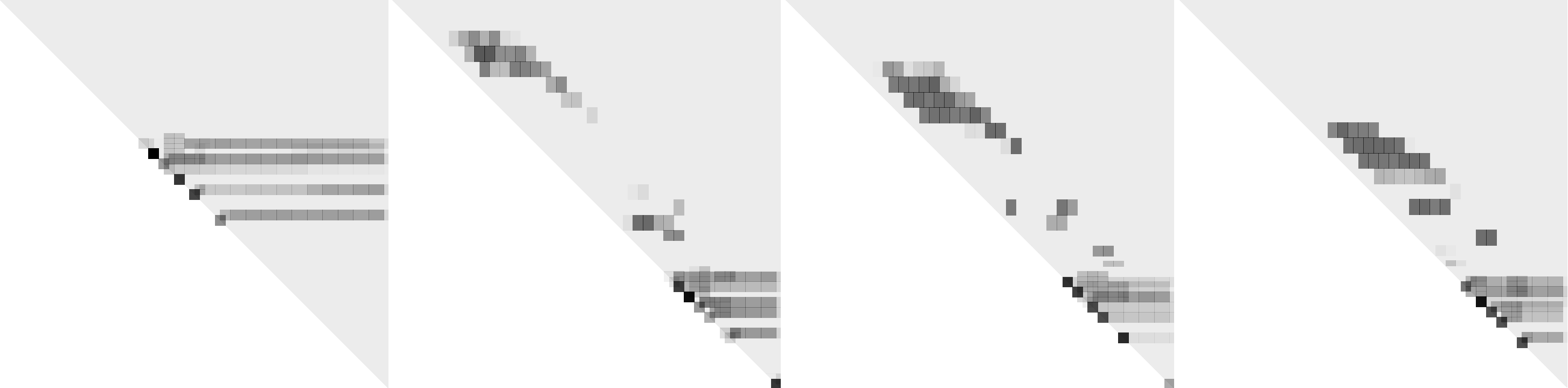}
 \caption{
 Four snapshots taken during a bulge chasing step.
 Note how the low-priority {\tt right update} tasks are delayed until the {\tt push bulges} tasks and the {\tt left bulges} tasks can no longer saturate all workers.
 The delayed {\tt right update} tasks form a blob that first appears in the second illustration from the left and then moves down just above the diagonal.
 }
 \label{fig:frames}
\end{figure}

\subsubsection{Comment regarding the task execution order}

It is important to realize that the task insertion order simply defines the dependencies between the tasks.
The runtime system is allowed to choose the task scheduling order as long as it respects the dependencies.
The priority of a task is taken into account once all its dependencies have been fulfilled.
The outcome of all these tasks insertion order and task priority considerations is shown in Fig. \ref{fig:frames}.
Note how the low-priority {\tt right update} tasks are delayed until the {\tt push bulges} and {\tt left bulges} tasks can no longer saturate all workers.
What is not shown in the figure is that the low-priority {\tt right update} tasks form a wave pattern that slowly travels from the diagonal in the direction of the upper-right corner of the matrix.

\subsubsection{Lowest-priority tasks}\label{subsec:lowest_prio}

\begin{figure}[h]
 \centering
 \includegraphics[scale=\figurescale]{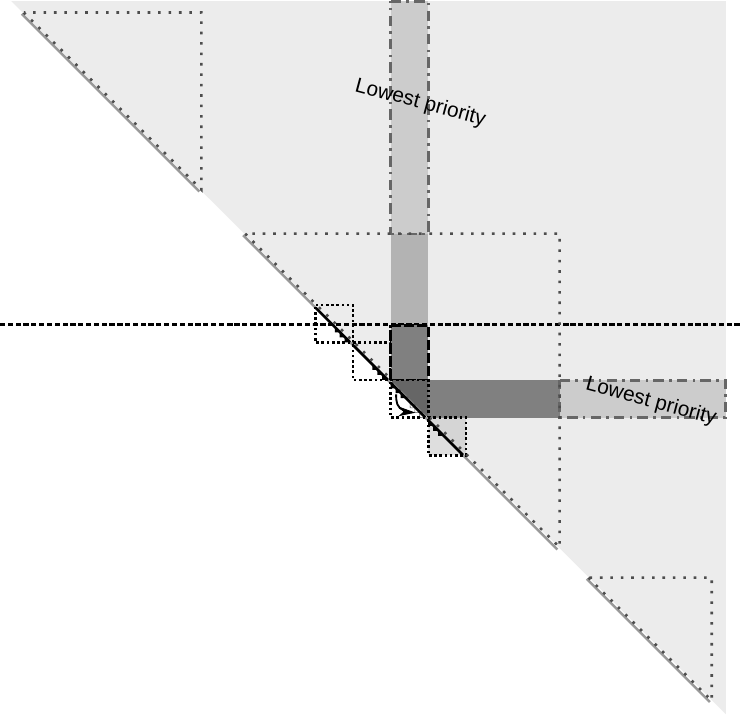}
 \caption{An illustration of the lowest priority update tasks. All updates outside the \emph{dashed triangle} are considered lowest priority as the associated tasks do not feed back to $\mathcal{G}_{qr}$.}
 \label{fig:low_prio}
\end{figure}

The fact that the unreduced blocks shrink as the algorithm progresses is also taken into account. 
In particular, as shown in Fig. \ref{fig:low_prio}, all update tasks that update sections of the matrix outside of the unreduced blocks are given the lowest priority ({\tt min\_prio}).
These tasks do not feed back to $\mathcal{G}_{qr}$ and can therefore be delayed.

\subsubsection{Preparation for the next AED step}

\begin{figure}[h]
 \centering
 \includegraphics[scale=\figurescale]{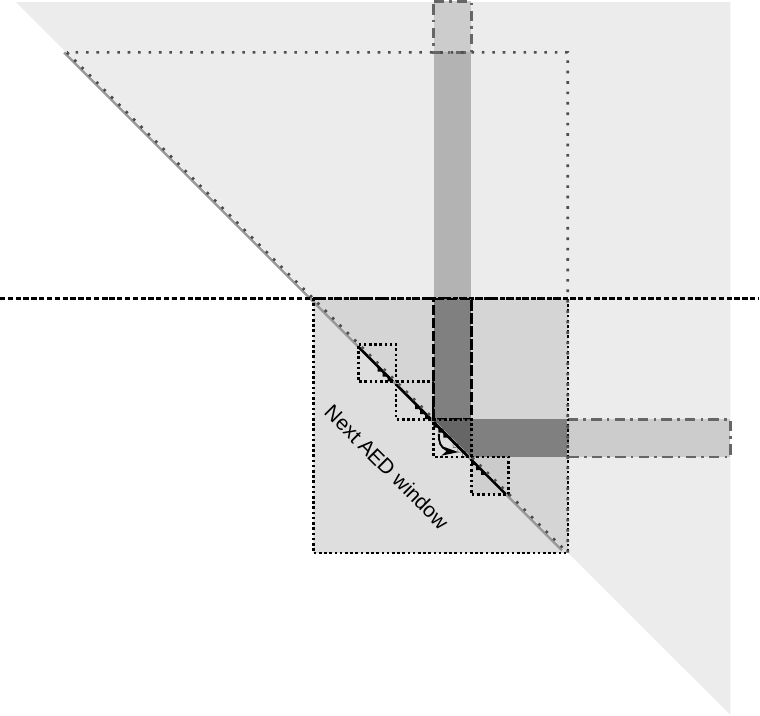}
 \caption{An illustration of the right-hand side update tasks that feed back to $\mathcal{G}_{qr}$. All tasks that overlap with the next AED window, i.e., are below the \emph{dashed line}, are given a higher priority.}
 \label{fig:aed_prio}
\end{figure}

Since most AED-related tasks are part of $\mathcal{G}_{qr}$, it is important to prioritize tasks that are direct or indirect dependencies to the AED-related tasks.
The task-based algorithm therefore calculates the size of the next AED window and effectively lifts the dashed line to the level of the upper edge of the next AED window as shown in Fig. \ref{fig:aed_prio}.
This means that {\tt right update} tasks, that are direct or indirect dependencies to the AED-related tasks, are given higher priority and are therefore scheduled earlier.
Note that Fig. \ref{fig:chain_flow2} has been simplified by omitting these higher priority {\tt right update} tasks from the sections highlighted with the dashed perimeters.

\subsubsection{Bringing everything together}

\begin{figure}[h]
 \centering
 \includegraphics[width=\textwidth]{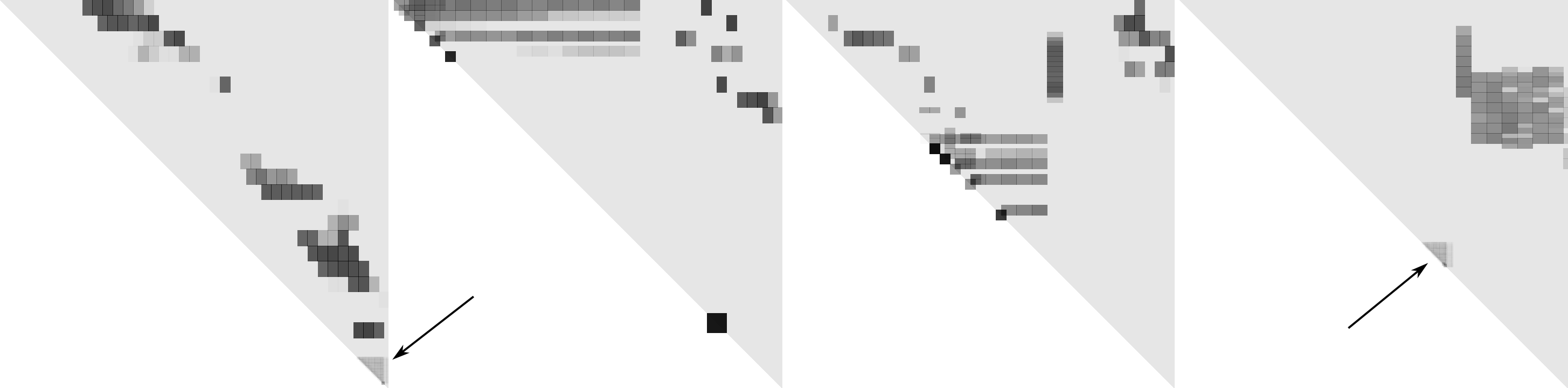}
 \caption{
 Four additional snapshots taken during computations.
 The arrows mark AEDs.
 From left to right:
(i) an AED step is being executed while other workers are still executing the low priority {\tt right update} tasks,
(ii) a {\tt small Hessenberg} tasks from an AED step is overlapped with the next bulge chasing step,
(iii) two bulge chasing steps (one appearing as a blob in the upper-right corner of the matrix) are overlapped with each other, and
(iv) an AED step is being executed while other workers are executing the lowest-priority {\tt left update} tasks.
 }
 \label{fig:frames2}
\end{figure}

The outcome of all these task insertion order and task priority considerations can be seen in Fig. \ref{fig:frames2}.
In particular, the figure demonstrates that (i) the next AED step can begin before all {\tt right update} tasks have have been completed, (ii) the next bulge chasing step can begin before the third AED sub-step has completed, (iii) two bulge chasing steps can be overlapped, and (iv) the low-priority {\tt left update} tasks are delayed until computational resources start becoming idle.
A complete illustration is available on YouTube, see \tcite{youtube_schur}.

\subsection{Parallel aggressive early deflation}\label{subsec:parallel_aed}

\subsubsection{Adaptive approach}\label{subsec:adaptive_approach}

\begin{figure}[h]
 \centering
 \includegraphics[scale=\figurescale]{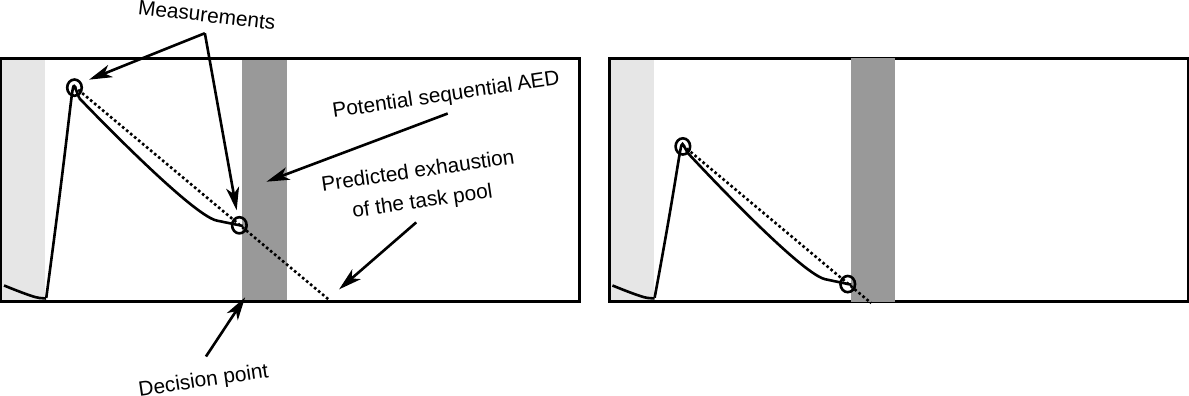}
 \caption{An illustration of the adaptive AED decisions.
 The vertical axis shows size of the task pool and the horizontal axis shows the wall time.
 The left figure shows a situation where the algorithm predicts that the task pool is not exhausted before a sequential AED finishes and the right figure shows a situation where the algorithm predicts that the tasks pool is exhausted.
 The later situation could thus benefit from a parallel AED.}
 \label{fig:adaptive}
\end{figure}

The task-based QR algorithm performs small AEDs sequentially and large AEDs in parallel.
The decision to perform a parallel AED is done adaptively.
The adaptive approach was first presented and evaluated by the author of this paper in \tcite{D65}.
The basic idea is visualized in Fig. \ref{fig:adaptive}.
The task-based algorithm measures size of the \emph{task pool} (i.e., the set of all inserted tasks that are not yet scheduled to workers) in two different points in time: 
(i) just after the bulge chasing tasks have been inserted and 
(ii) just before the task-based algorithm must decide whether to perform a sequential AED or a parallel AED.
The bulge chasing tasks are consumed at a constant rate and the task-based algorithm can therefore use a linear model to predict the point in time when the task pool will be exhausted.
The StarPU runtime system provides an estimate for the execution time of a sequential AED in the form of a performance model $a n^b + c$, where $n$ is the size of the AED window and $a,b,c$ are model parameters.
With this information, the task-based algorithm can predict if a sequential AED finishes before the task pool is exhausted.
If this is not the case, then the task-based algorithm performs a parallel AED.
The smallest and largest AEDs are always performed sequentially and in parallel, respectively.
Note that these predictions are not exact and the adaptive approach is meant to be used as a heuristic.

\subsubsection{Tasks, state transitions and eigenvalue reordering}

The first AED sub-step (recursive QR) is performed by copying the AED window to a separate matrix that has a smaller tile size ($b_{aed}$) than the main matrix.
The parent segment retains the {\bf AED reduce} state until the AED window has been reduced to real Schur form and then transitions to the {\bf AED deflate} state.
The second AED sub-step (deflation check) is performed by a set of {\tt deflate} tasks.
A {\tt deflate} task has two main functions: 
(i) perform deflation checks for the eigenvalue candidates that fall within a diagonal \emph{deflation window} and 
(ii) reorder failed eigenvalue candidates to the upper-left corner of the deflation window.
Once the first AED sub-step has finished, the first {\tt deflate} task is inserted such that the lower-right corner of the corresponding deflation window is placed flush against the lower-right corner of the AED window.
When the scan later returns to the corresponding segment, the main thread acquires the outcome of the {\tt deflate} task.
The {\tt deflate} tasks are given the highest priority ({\tt max\_prio}), the {\tt right update} tasks the second highest priority ({\tt max(def\_prio, max\_prio - 1)}) and the {\tt left update} tasks the default priority ({\tt def\_prio}).
Those {\tt right update} tasks that update the accumulator matrix are given the lowest priority ({\tt min\_prio}) which is in this context higher than the lower priority allowed by the runtime system.

\begin{figure}[h]
 \centering
 \includegraphics[scale=\figurescale]{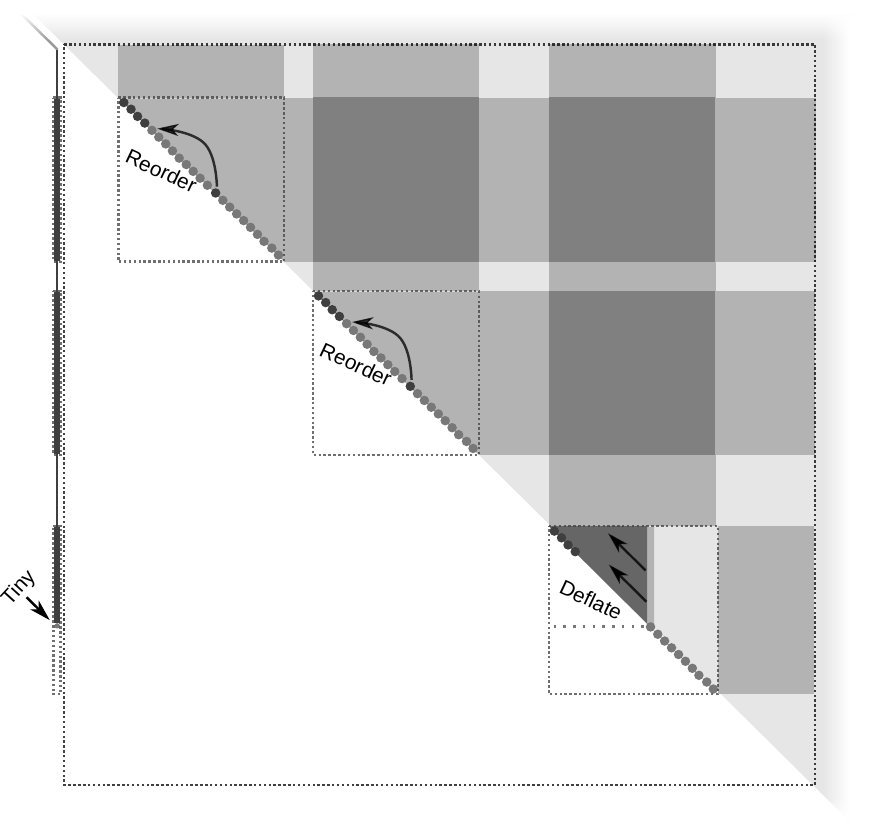}
 \caption{An illustration of the second AED sub-step with a deflation window and two reordering windows being processed in parallel.
 Each {\tt deflate} task takes the deflation/reordering window and the corresponding section of the spike as arguments.}
 \label{fig:parallel_aed}
\end{figure}

If the previous deflation window contains less than $b_{aed}$ failed eigenvalue candidates, then the next {\tt deflate} task is inserted such that the lower-right corner of the corresponding deflation window is placed flush against the bottommost failed eigenvalue candidate.
If, on the other hand, the previous deflation window contains more than $b_{aed}$ failed eigenvalue candidates, then one or two reordering window chains are inserted. 
The eigenvalue reordering is performed by {\tt deflate} tasks and each reordering window chain moves at most $b_{aed}-1$ failed eigenvalue candidates to the upper-left corner of the AED window.
The next {\tt deflate} task is then inserted such that the lower-right corner of the corresponding deflation window is placed flush against the next unevaluated eigenvalue candidate.
The process repeats until all eigenvalue candidates have been evaluated.  
Note that the deflation and reordering related {\tt deflate} tasks can be executed concurrently as visualized in Fig. \ref{fig:parallel_aed}.

\subsubsection{Failed swaps}

One of the major complications when it comes to parallelizing the AED step is the fact that the eigenvalue reordering process can fail.
That is, the kernels that are used to swap two adjacent diagonal blocks can return an error code indicating that the swapping operation is too ill-conditioned.
The reason why this becomes a problem is that these failures are difficult to predict and the task-based algorithm should, in principle, investigate the outcome of each eigenvalue reordering related {\tt deflate} task.
Otherwise, the main thread may lose track of the locations of the failed eigenvalue candidates and therefore cannot place the deflation and reordering windows without splitting 2-by-2 blocks.
Acquiring the outcome of each {\tt deflate} task would, however, limit the parallelism since the main thread could potentially be blocked each time it acquires a data handle.

\subsubsection{Greedy approach}\label{subsec:greedy_approach}

In order to solve this problem with failed swaps, the task-based algorithm adopts a \emph{greedy approach} where the task insertion phase assumes that all the swaps will be successful and the {\tt deflate} tasks make sure that the Schur form is retained even if the swapping kernels fail to swap two or more diagonal blocks.
This approach can lead to situations where only a subset of the eigenvalue candidates are evaluated but such situations are generally assumed to be quite uncommon. 
Therefore, an approach that favors performance, like the one described here, is preferred.

\begin{figure}[h]
 \centering
 \includegraphics[scale=\figurescale]{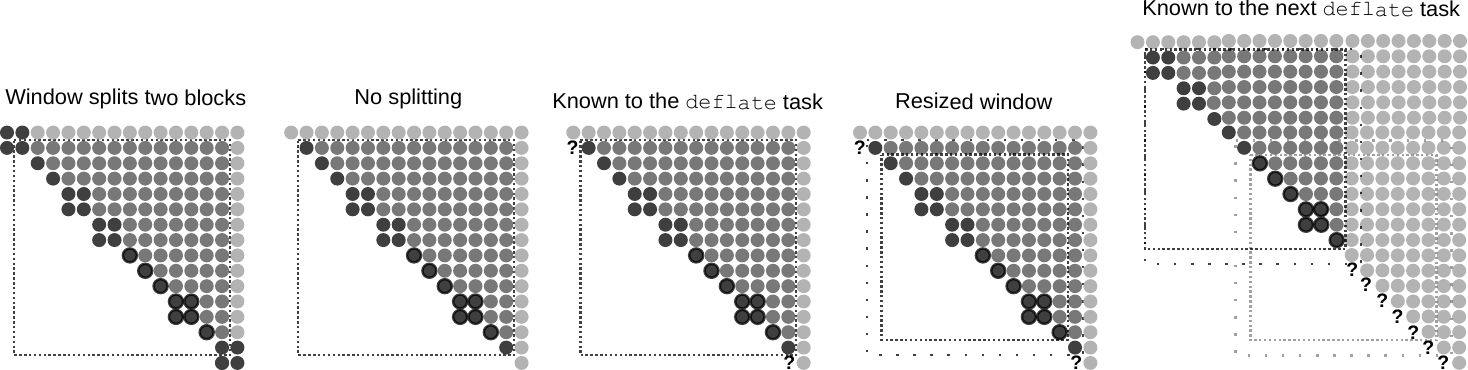}
 \caption{Illustrations of deflation and reordering windows.
 From left to right: 
 (i) a situation where the window boundary splits two 2-by-2 blocks, 
 (ii) a situation where the window boundary does not split any 2-by-2 blocks, 
 (iii) matrix entries that are known to a {\tt deflate} task, 
 (iv) resized active window, and
 (v) matrix entries that are known to the next {\tt deflate} task.}
 \label{fig:offset}
\end{figure}

The actual details of this greedy approach are visualized in Fig. \ref{fig:offset}.
Since a {\tt deflate} task cannot separate between situations where the deflation/reordering window splits a 2-by-2 block (leftmost figure) and when it does not (second figure from the left), the {\tt deflate} task will resize its active computational window (fourth figure from the left) when necessary.
The resizing happens when the topmost diagonal block in the window is a 1-by-1 block and/or when the bottommost diagonal block in the window is a 1-by-1 block.
The fact that the topmost failed eigenvalue candidate may end up being moved to the second position from the top is taken into account when the next reordering window is placed.
If a failure to swap occurs, then the {\tt deflate} tasks will simply halt the reordering.
The locations of the topmost and bottommost failed eigenvalue candidates are stored in a separate data handle and passed on to the next {\tt deflate} task in the reordering window chain.
If all failed eigenvalue candidates are found to be outside the reordering window, then the eigenvalue reordering is skipped.

\section{Software}\label{seq:software}

The new StarPU-based implementation is part of the open-source StarNEig library, see \citeauthor{myllykoski_2020} \citeyearpar{myllykoski_2020, ppam2019mm}.
The details on how to use the library, and the task-based QR algorithm, are covered in the StarNEig User's Guide that is available from the StarNEig home page (\tcite{starneig_website}) in both PDF and HTML formats. 

As explained in Subsection \ref{subsection:distr_blocks}, StarNEig implements an additional layer of structure on top of the tiling.
The tiles are consolidated into disjoint rectangular blocks of uniform size (excluding the last block row and block column).
All tiles that belong to the same distributed block must have the same owner but otherwise the data distribution can be arbitrary.
For optimal performance, the distributed blocks should be relatively large (at least 1000-by-1000 but preferably much larger).
This is due to the fact that StarNEig further divides the distributed blocks into tiles and a tiny tile size leads to excessive task scheduling overhead.
Furthermore, StarNEig should be used in one process per node (1ppn) topology\footnote{A node can be an entire compute node, a NUMA island or some other reasonably large collection of CPU cores.}.
StarNEig implements a set of interface functions for manipulating the distributed matrices.
StarNEig also implements a ScaLAPACK compatibility layer which allows a user to use BLACS contexts and descriptors (see BLACS's User's Guide by \tcite{blacs}) with StarNEig.

\section{Computational results}\label{sec:results}

This section demonstrates the performance of the StarPU-based implementation (StarNEig) with sets of computational experiments.
The StarPU-based implementation is compared against LAPACK (\texttt{DHSEQR}) and ScaLAPACK (\texttt{PDHSEQR}) in both single-node (shared/distributed memory) and multi-node (distributed memory) environments.
The GPU performance and the scalability of the StarPU-based implementation are also demonstrated.

\begin{remark}
Only the standard case is discussed in this section.
See \tcite{D27} for preliminary results relating to the computation of generalized Schur form using the task-based QZ algorithm.
Note that the problems relating to the detection of infinite eigenvalues have been fixed in StarNEig version 0.1-beta.4.
\end{remark}

\subsection{Hardware environment}

All computational experiments were performed using two systems, Kebnekaise and Abisko, located at the High Performance Computing Center North (HPC2N), Ume{\aa} University, Sweden.
Kebnekaise is a heterogeneous system consisting of many different types of Intel and Nvidia based nodes.
The node types that are used for this paper are the following:
\begin{description}
 \item[Broadwell node] contains 28 Intel Xeon E5-2690v4 cores organized into 2 NUMA islands with 14 cores each and 128 GB memory. 
 The nodes are connected with FDR Infiniband. 
 All multi-node experiments were performed on these nodes.
 \item[Skylake node] contains 28 Intel Xeon Gold 6132 cores organized into 2 NUMA islands with 14 cores each and 192 GB memory. 
 Most single-node experiments were performed on these nodes.
 \item[V100 GPU node] contains 28 Intel Xeon Gold 6132 cores organized into 2 NUMA islands with 14 cores each and 192 GB memory. 
 Each NUMA island is connected to a single Nvidia Tesla V100 GPU with 16 GB memory.
 All GPU experiments were performed on these nodes.
\end{description}
Abisko is an older system based on AMD Interlagos family of CPUs:
\begin{description}
 \item[Abisko node] contains 48 AMD Opteron 6238 cores organized into 8 NUMA islands (4 sockets) with 6 cores each and 128 GB memory.
 Two neighboring cores share a floating point unit (FPU), i.e., the total number of FPUs per node is 24.
 The nodes are connected with Mellanox 4X QSFP 40 Gb/s InfiniBand.
\end{description}

\subsection{Software environment and configuration}\label{subsec:config}

\subsubsection{Kebnekaise configuration}
The software was compiled with GCC 8.3.0 and linked to OpenMPI 3.1.4, OpenBLAS 0.3.7, ScaLAPACK 2.0.2, CUDA 10.1.243, StarPU 1.3.3 and StarNEig 0.1.1.
For GPU and Hessenberg reduction experiments, the software was linked to StarPU 1.2.10 (with \texttt{-{}-disable-cuda-memcpy-peer}) and StarNEig 0.1.3 as the StarPU 1.3 series performs worse than the StarPU 1.2 series in these particular contexts and StarNEig 0.1.3 contains modifications to the Hessenberg reduction phase.
Only 25 cores were used in most single-node experiments because this choice leads a square 5-by-5 MPI process grid.
A square MPI process grid usually leads to better performance with the \texttt{PDHSEQR} routine because for a $pm$-by-$pn$ MPI process grid, the maximum number of concurrent near-diagonal bulge chasing windows is given by $\min(pm,pn)$. 
If all 28 cores were used, then the maximum number of concurrent near-diagonal bulge chasing windows would have been $\min(7,4) = 4$.

\subsubsection{Abisko configuration}
The software was compiled with GCC 8.2.0 and linked to OpenMPI 3.1.3, OpenBLAS 0.3.5, ScaLAPACK 2.0.2, StarPU 1.3.3 and StarNEig 0.1.1.
Since two neighboring cores share a FPU, only every other core, i.e., a total of 24 cores per node, were used in the experiments.
With \texttt{PDHSEQR}, the maximum number of concurrent near-diagonal bulge chasing windows was therefore $\min(6,4) = 4$.

\subsubsection{LAPACK configuration}
The \texttt{DHSEQR} routine was parallelized by enabling the OpenBLAS multi-threading support.
This means that each individual off-diagonal BLAS-3 update is performed in parallel.

\subsubsection{ScaLAPACK configuration}\label{subsection:scalapack_conf}
The matrices were distributed in 160-by-160 blocks.
Initial experiments indicated that smaller block sizes lead to decreased performance and that larger block sizes do not provide any significant performance benefit.
ScaLAPACK was configured to use one process per core (1ppc) topology.
Note that the version of the \texttt{PDHSEQR} routine that is provided with ScaLAPACK 2.0.2 is known to be buggy and StarNEig was therefore compared against an updated version of the routine, see \citeauthor{GraKagKreShao2015a} \citeyearpar{GraKagKreShao2015a, GraKagKreShao2015b}.
In particular, the implementation that exists in ScaLAPACK 2.0.2 does not function reliably when the distributed block size is larger than 50-by-50.

\subsubsection{StarNEig configuration}\label{subsection:starneig_conf}
One core was allocated for inserting the tasks into StarPU, i.e., the number of StarPU workers (including the MPI worker) was set to $\max(1, p-1)$, where $p$ is the number of available cores on a node.
StarNEig was configured to use a library defined default tile size that depends on the matrix size and the CPU core count.
In multi-node experiments, StarNEig was configured to use a library defined default distributed block size and a one process per node (1ppn) topology.
The matrices were distributed using a two-dimensional block cyclic distribution.
All computational experiments were conducted using a LAPACK-style deflation condition (see Subsection \ref{subsec:deflation_cond}).

\subsubsection{Test driver configuration}
The StarNEig test driver random number seed was set to 2020.
For most experiments, only one run was performed for each measurement.
This was done for the following four reasons:
(i) The access to the Kebnekaise and Abisko systems is limited and the number of performed experiments is quite large (and not all of them are even included in this paper).
Also note that the validation of the result in most cases took far longer than the actual computation.
(ii) Most measurements took a few minutes or more, and are therefore not susceptible to minor interference from the operating system and other external sources.
Furthermore, all experiments were performed with exclusive access to the nodes.
(iii) None of the conclusions of the paper depend on individual measurements. 
(iv) In previous studies by the author (see \tcite{Myllykoski2018} and \tcite{NLAFET-WN11}), the coefficient of variation has been at most a few percent, usually around 1-2\%.
When the results of the experiments are inspected, it becomes clear that the observed trends are consistent across all test problems and similar on both machines.
The GPU-experiments are generally more noisy and for those experiments the median of three runs is reported.
In addition, the performance models were pre-calibrated with \texttt{STARPU\_CALIBRATE=1} and \texttt{STARPU\_SCHED=random} as this leads to better calibrated performance models and thus a more consistent performance.

\subsection{Test matrices}

\begin{table}[h]
\centering
\caption{Sparse test matrices selected from the SuiteSparse Matrix Collection.}
\label{tab:sparse_matrices}
\begin{tabular}{r|r|rr|l}
Name & \multicolumn{1}{c|}{$n$} & \multicolumn{2}{c|}{Nonzeros} & Described source \\ 
\hline 
g7jac020		&  5\,850	& 42\,568	& 0.12 \%	& Economic Problem \\
sinc15		& 11\,532	& 551\,184		& 0.41 \%	& Materials Problem \\
ex11			& 16\,614	& 1\,096\,948	& 0.40 \%	& Computational Fluid Dynamics Problem \\
ns3Da		& 20\,414	& 1\,679\,599	& 0.40 \%	& Computational Fluid Dynamics Problem \\
TSOPF\_RS\_b2052\_c1 & 25\,626 & 6\,761\,100	& 1.03 \%	& Power Network Problem \\
invextr1\_new	 & 30\,412	& 1\,793\,881	& 0.19 \%	& Computational Fluid Dynamics Problem \\
g7jac120sc	& 35\,550	& 412\,306		& 0.03 \%	& Economic Problem \\
av41092		& 41\,092	& 1\,683\,902	& 0.10 \%	& 2D/3D Problem \\
\end{tabular} 
\end{table}

Three different classes of test matrices were considered:
\begin{enumerate}

 \item \textit{Sparse real-world matrices.} A set of sparse nonsymmetric matrices was selected from the SuiteSparse Matrix Collection (formerly known as Tim Davis's collection), see Table \ref{tab:sparse_matrices}. 
 These test matrices were included because the behavior of the QR algorithm is sensitive to the properties of the matrix, such as, the distribution of the eigenvalues. 
 It is therefore vital that performance of StarNEig is demonstrated with real-world matrices.
 Using a set of dense real-world matrices would have been a more obvious choice but a database that is comparable to the SuiteSparse Matrix Collection simply does not exist for dense matrices.
 The sparse matrices are also more easily accessible for anyone due to their small storage footprint.
 
 \item \textit{Synthetic matrices with known eigenvalues (syn\_$n$).} These matrices are generated by first forming a random upper triangular matrix $\bar T \in \mathbb{R}^{n \times n}$ (entries uniformly distributed on the interval $[-1,1]$).
 The known eigenvalues, $\lambda_1, \dots, \lambda_n$, 50\% of them being complex conjugate pairs, are then placed on the diagonal of $\bar T$ to form a Schur matrix $\bar S$.
 The Schur matrix $\bar S$ is then multiplied from both sides with a random Householder reflector $\bar Q \in \mathbb{R}^{n \times n}$ to generate a dense matrix $A = \bar Q \bar S \bar Q^T$.
 The known eigenvalues are initially selected to be $\pm 1, \pm 3, \pm 5, \dots$.
 The complex conjugate pairs of eigenvalues, $(\lambda_{i}, \lambda_{i+1})$, are then randomly selected and lifted from the real line by setting
 \begin{align}
 \begin{split}
  \lambda_{i} &\gets \text{Re}(\lambda_{i}) + i \, |\text{Re}(\lambda_{i})| \\
  \lambda_{i+1} &\gets \text{Re}(\lambda_{i}) - i \, |\text{Re}(\lambda_{i})|.
 \end{split}
 \end{align}  
 
 \item \textit{Hessrand matrices (hess\_$n$).} Let $\mathcal{N}(\mu,\sigma)$ be the normal distribution with mean $\mu$ and variance $\sigma$, and let $\chi^2(\tau)$ be the chi-squared distribution with $\tau$ degrees of freedom. 
As done in \tcite{Braman2002}, the upper Hessenberg matrix $H \in \mathbb{R}^{n \times n}$ is generated directly as follows:
 \begin{align}
 \begin{split}
  h_{i,j}     &\sim \mathcal{N}(0,1), \; i = 1, \dots, j, \; j = 1, \dots, n, \\
h_{i+1,i}^2 &\sim \chi^2(n-i), \; i = 1, \dots, n-1.
 \end{split}
\end{align} 

\end{enumerate}

\subsection{Accuracy metrics}

Given a computed orthogonal similarity transformation $A \approx U X U^T$, we measure the backward error with the following relative residual:
\begin{align}
R_{A}(X,U) = \frac{ || U X U^T - A ||_F }{ u || A ||_F },\label{eq:RsepA}
\end{align}
where $|| \cdot ||_F$ is the Frobenius norm and $u$ is the unit roundoff ($u = 2^{-52} \approx 2.22 \cdot 10^{-16}$), see \tcite{doi:10.1137/S0895479896313188}.
Similarly, we measure the loss of orthogonality with the relative residual
\begin{align}
R_{orth}(U) = \frac{ || U U^T   - I ||_F }{ u || I ||_F }.\label{eq:Rorth}
\end{align}
Note that the residuals are always computed with respect to the original matrix $A$.
Finally, given a computed Schur decomposition $A \approx Q \bar S Q^T$, we measure for each computed eigenvalue $\bar \lambda \in \lambda(\bar S)$ the relative error
\begin{align}
E_{\lambda}(\bar \lambda) = \min_{\lambda \in \lambda(A)} \frac{|\bar \lambda - \lambda|}{u|\lambda|}.
\end{align}
We report both the mean error and the maximum error:
\begin{align}
 E_{\lambda, mean} = \frac{\sum_{\bar \lambda \in \lambda(\bar S)} E_{\lambda}(\bar \lambda)}{n}
 \; \text{ and } \;
 E_{\lambda, max} = \max_{\bar \lambda \in \lambda(\bar S)} E_{\lambda}(\bar \lambda).
\end{align}

\subsection{Comment regarding the Hessenberg reduction phase}

\begin{table}[h]
\centering
\caption{A comparison between a Skylake node (25 cores) and a single socket of a V100 GPU node (14 cores + GPU) when computing a Hessenberg form, and the associated residuals from the former set of experiments.
The missing data points are due to the fact that the matrices simply did not fit into the GPU's memory. }
\label{tab:keb_hess_sparce}
\begin{tabular}{r | c c | c | c }
       & \multicolumn{2}{c|}{Run time [s]} & & \\
Matrix & CPU only & CPU + GPU & $R_{A}(H,Q_1)$ & $R_{orth}(Q_1)$ \\
\hline
g7jac020 & 10 & 2 & 11 & 10 \\
sinc15 & 56 & 13 & 10 & 10 \\
ex11 & 150 & 32 & 9 & 10 \\
ns3Da & 266 & 55 & 20 & 10 \\
TSOPF\_RS\_.. & 507 & 105 & 10 & 11 \\
invextr1\_.. & 814 & 167 & 16 & 10 \\
g7jac120s.. & 1268 & -- & 10 & 10 \\
av41092 & 1923 & -- & 25 & 11
\end{tabular}
\end{table}

The so-called standard algorithm by \tcite{Quintana-Orti2006} for reducing a nonsymmetric matrix $A$ to upper Hessenberg form $H$ is known to be memory bound.
The Hessenberg reduction phase consequently requires more computing time than the Schur reduction phase. 
However, as shown in Table \ref{tab:keb_hess_sparce}, the Hessenberg reduction phase can be accelerated with a GPU.
This means that the Hessenberg reduction phase can actually be performed faster than the Schur reduction phase and any improvements made to the Schur reduction phase are therefore meaningful.
StarNEig (and MAGMA, see \tcite{tomov2009accelerating}) currently support only a single GPU but a multi-GPU support is being developed while this paper is being prepared.
Note that StarNEig would not benefit from the additional 14 cores from the second socket event though some BLAS-3 tasks are executed on a CPU (see \tcite{myllykoski_2020}).
This is because the critical path is executed on the GPU and the 14 cores from a single socket are able to execute the remaining BLAS-3 tasks at sufficiently fast rate to not cause a bottleneck.

\subsection{Comment regarding the deflation conditions and the accuracy}\label{subsec:deflation_cond}

Before discussing the performance, we must first discuss the deflation condition and how it affects the accuracy.
Given a sub-matrix than encloses a AED window and the corresponding spike,
\begin{align}
 \left[
 \begin{array}{c|cccccc}
  h_{i, i} & \ddots & \ddots & \ddots & \ddots \\ \hline
  h_{i+1, i} & \ddots & \ddots & \ddots & \ddots \\
  \vdots    &  \ddots      & \ddots & \ddots & \ddots \\
  h_{i+k-1,i} & & h_{i+k-1,   i+k-1} & h_{i+k-1,   i+k} & \ddots \\
  h_{i+k,i} & & h_{i+k, i+k-1} & h_{i+k, i+k} & \ddots \\
  0      & & & & \ddots
 \end{array}
 \right],
\end{align}
the deflation condition determines when the spike entries $h_{i+k-1,i}$ and/or $h_{i+k,i}$ are small enough to be set to zero without introducing significant perturbations. 
The StarNEig supports three different deflation and vigilant deflation conditions:
\begin{enumerate}

 \item \textit{LAPACK-style conditions (used in this paper).} 
 If $h_{i+k,i+k-1} = 0$ and 
 \begin{align}
  |h_{i+k,i}| \leq \max(\nu (n/u), u |h_{i+k, i+k}|),
 \end{align}
 where $\nu$ is the smallest floating-point number such that $1 / \nu$ does not overflow and $u$ is the unit roundoff, then the 1-by-1 block $[h_{i+k, i+k}]$ is deflated by setting $h_{i+k,i}$ to zero. 
 If $h_{i+k,i+k-1} \neq 0$ and 
 \begin{align}
  \max(|h_{i+k,i}|,|h_{i+k-1,i}|) \leq \max(\nu (n/u), u \hat h_{i+k}),
 \end{align}
 where $\hat h_{j} = \sqrt{|h_{j, j}h_{j-1, j-1}|} + \sqrt{|h_{j-1, j}h_{j, j-1}|}$, then the 2-by-2 block
 \begin{align*}
  \begin{bmatrix}
   h_{i+k-1, i+k-1} & h_{i+k-1, i+k}\\ 
   h_{i+k, i+k-1} & h_{i+k, i+k}
  \end{bmatrix}
 \end{align*}
 is deflated by setting both $h_{i+k-1,i}$ and $h_{i+k,i}$ to zero.
 StarNEig mirrors LAPACK 3.9.0 when it comes to implementation of vigilant deflation checks.
  
 \item \textit{Norm-stable conditions (default).} 
 If $h_{i+k,i+k-1} = 0$ and $|h_{i+k,i}| \leq u\|H\|_F$, then the 1-by-1 block $[h_{i+k, i+k}]$ is deflated by setting $h_{i+k,i}$ to zero.
 If $h_{i+k,i+k-1} \neq 0$, $|h_{i+k,i}| \leq u\|H\|_F$ and $|h_{i+k-1,i}| \leq u\|H\|_F$, then the 2-by-2 block
 \begin{align*}
  \begin{bmatrix}
   h_{i+k-1, i+k-1} & h_{i+k-1, i+k}\\ 
   h_{i+k, i+k-1} & h_{i+k, i+k}
  \end{bmatrix}
 \end{align*}
 is deflated by setting both $h_{i+k-1,i}$ and $h_{i+k,i}$ to zero (see \tcite{Braman2002a}).
 Given a sub-diagonal entry $h_{i,i-1}$, a vigilant deflation occurs if $|h_{i,i-1}| \leq u\|H\|_F$.  
 
 \item \textit{Fixed conditions.} 
 If $h_{i+k,i+k-1} = 0$ and $|h_{i+k,i}| \leq \epsilon_{thold}$, where $\epsilon_{thold}$ is a user defined threshold, then the 1-by-1 block $[h_{i+k, i+k}]$ is deflated by setting $h_{i+k,i}$ to zero.
 If $h_{i+k,i+k-1} \neq 0$, $|h_{i+k,i}| \leq \epsilon_{thold}$ and $|h_{i+k-1,i}| \leq \epsilon_{thold}$, then the 2-by-2 block
 \begin{align*}
  \begin{bmatrix}
   h_{i+k-1, i+k-1} & h_{i+k-1, i+k}\\ 
   h_{i+k, i+k-1} & h_{i+k, i+k}
  \end{bmatrix}
 \end{align*}
 is deflated  by setting both $h_{i+k-1,i}$ and $h_{i+k,i}$ to zero.
 Given a sub-diagonal entry $h_{i,i-1}$, a vigilant deflation occurs if $|h_{i,i-1}| \leq \epsilon_{thold}$.

\end{enumerate}

\begin{table}[h]
\centering
\caption{A comparison between the LAPACK-style deflation condition (left in each column-pair) and the norm-stable deflation condition (right in each column-pair) on a Skylake node (25 cores) when computing a Schur form from a Hessenberg form.}
\label{tab:keb_schur_sparce_norm}
\begin{tabular}{r | c c | c c | c c | c c | c c}
Matrix & \multicolumn{2}{c|}{Run time [s]} & \multicolumn{2}{c|}{$R_{A}(S,Q)$} & \multicolumn{2}{c|}{$R_{orth}(Q)$} & \multicolumn{2}{c|}{$E_{\lambda, mean}$} & \multicolumn{2}{c}{$E_{\lambda, max}$}\\
\hline
syn\_10000 & 14 & 13 & 238 & 267 & 157 & 160 & 62 & 65 & 505 & 504 \\
syn\_20000 & 68 & 63 & 336 & 356 & 237 & 223 & 92 & 91 & 1488 & 912 \\
syn\_40000 & 473 & 427 & 619 & 549 & 412 & 349 & 181 & 150 & 3462 & 1780 \\
\hline
g7jac020 & 3 & 3 & 112 & 88 & 93 & 67 & -- & -- & -- & -- \\
sinc15 & 8 & 5 & 103 & 87 & 83 & 65 & -- & -- & -- & -- \\
ex11 & 43 & 36 & 633 & 612 & 145 & 130 & -- & -- & -- & -- \\
ns3Da & 54 & 44 & 231 & 205 & 147 & 130 & -- & -- & -- & -- \\
TSOPF\_RS\_.. & 223 & 135 & 472 & 372 & 236 & 188 & -- & -- & -- & -- \\
invextr1\_.. & 410 & 304 & 1393 & 1011 & 590 & 352 & -- & -- & -- & -- \\
g7jac120s.. & 277 & 186 & 242 & 165 & 240 & 147 & -- & -- & -- & -- \\
av41092 & 478 & 204 & 418 & 277 & 231 & 176 & -- & -- & -- & --
\end{tabular}
\end{table}

Table \ref{tab:keb_schur_sparce_norm} shows how StarNEig behaves when we try both the LAPACK-style deflation condition and the norm-stable deflation condition.
It is clear that the choice of deflation condition impacts both the performance and the accuracy.
In general, the norm-stable deflation condition leads to a shorter run time, which is not surprising since the norm-stable deflation condition is less strict than the LAPACK-style deflation condition.
However, what might at first appear surprising is the fact that the norm-stable deflation condition can actually lead to a more accurate solution as shown in Table \ref{tab:keb_schur_sparce_norm}.
The differences are small and likely contributable to the rounding errors resulting from the additional QR iterations due to the stricter deflation condition.

Unless otherwise specified, StarNEig defaults to the norm-stable deflation condition.
However, since the \texttt{DHSEQR} and \texttt{PDHSEQR} routines use the LAPACK-style deflation condition, StarNEig was configured to use the LAPACK-style conditions in all computational experiments presented in this paper.
The choice of the deflation condition is not within the intended scope of this paper, and the use of the same deflation condition makes it easier to separate the algorithmic and implementation improvements from the improvement that is purely due to the choice of the deflation condition.

\subsection{Single-node performance comparison}

\begin{table}[h]
\centering
\caption{A comparison between LAPACK (left in each column-triplet), ScaLAPACK (middle in each column-triplet) and StarNEig (right in each column-triplet) on a Skylake node (25 cores) when computing a Schur form from a Hessenberg form.}
\label{tab:keb_schur_sparce}
\begin{tabular}{r | c c c | c c c | c c c }
Matrix & \multicolumn{3}{c|}{Run time [s]} & \multicolumn{3}{c|}{$R_{A}(S,Q)$} & \multicolumn{3}{c}{$R_{orth}(Q)$} \\
\hline
syn\_10000 & 62 & 37 & \textbf{ 14 } & 276 & 174 & \textbf{ 238 } & 168 & 113 & \textbf{ 157 } \\
syn\_20000 & 325 & 198 & \textbf{ 68 } & 333 & 289 & \textbf{ 336 } & 202 & 192 & \textbf{ 237 } \\
syn\_40000 & 2052 & 2995 & \textbf{ 473 } & 492 & 619 & \textbf{ 619 } & 263 & 396 & \textbf{ 412 } \\
\hline
g7jac020 & 11 & 8 & \textbf{ 3 } & 87 & 84 & \textbf{ 112 } & 69 & 65 & \textbf{ 93 } \\
sinc15 & 27 & 21 & \textbf{ 8 } & 62 & 58 & \textbf{ 103 } & 89 & 66 & \textbf{ 83 } \\
ex11 & 161 & 104 & \textbf{ 43 } & 155 & 176 & \textbf{ 633 } & 101 & 105 & \textbf{ 145 } \\
ns3Da & 222 & 157 & \textbf{ 54 } & 174 & 171 & \textbf{ 231 } & 120 & 117 & \textbf{ 147 } \\
TSOPF\_RS\_.. & 864 & 465 & \textbf{ 223 } & 65 & 100 & \textbf{ 472 } & 171 & 171 & \textbf{ 236 } \\
invextr1\_.. & 1077 & 644 & \textbf{ 410 } & 443 & 510 & \textbf{ 1393 } & 167 & 174 & \textbf{ 590 } \\
g7jac120s.. & 784 & 653 & \textbf{ 277 } & 103 & 117 & \textbf{ 242 } & 123 & 137 & \textbf{ 240 } \\
av41092 & 1486 & 893 & \textbf{ 478 } & 222 & 229 & \textbf{ 418 } & 162 & 168 & \textbf{ 231 }
\end{tabular}
\end{table}

\begin{table}[h]
\centering
\caption{A comparison between LAPACK (left in each column-triplet), ScaLAPACK (middle in each column-triplet) and StarNEig (right in each column-triplet) on an Abisko node (24 cores/FPUs) when computing a Schur form from a Hessenberg form.}
\label{tab:abi_schur_sparce}
\begin{tabular}{r | c c c | c c c | c c c}
Matrix & \multicolumn{3}{c|}{Run time [s]} & \multicolumn{3}{c|}{$R_{A}(S,Q)$} & \multicolumn{3}{c}{$R_{orth}(Q)$} \\
\hline
syn\_10000 & 186 & 108 & \textbf{ 32 } & 283 & 228 & \textbf{ 230 } & 174 & 144 & \textbf{ 153 } \\
syn\_20000 & 1082 & 650 & \textbf{ 168 } & 340 & 422 & \textbf{ 317 } & 206 & 295 & \textbf{ 224 } \\
syn\_40000 & 7181 & 5148 & \textbf{ 1229 } & 490 & 982 & \textbf{ 653 } & 260 & 746 & \textbf{ 419 } \\
\hline
g7jac020 & 38 & 19 & \textbf{ 8 } & 88 & 83 & \textbf{ 111 } & 69 & 64 & \textbf{ 91 } \\
sinc15 & 130 & 51 & \textbf{ 19 } & 57 & 60 & \textbf{ 107 } & 69 & 71 & \textbf{ 80 } \\
ex11 & 563 & 340 & \textbf{ 112 } & 159 & 122 & \textbf{ 641 } & 100 & 159 & \textbf{ 150 } \\
ns3Da & 752 & 521 & \textbf{ 146 } & 172 & 205 & \textbf{ 235 } & 118 & 147 & \textbf{ 150 } \\
TSOPF\_RS\_.. & 3108 & 2050 & \textbf{ 610 } & 62 & 68 & \textbf{ 468 } & 172 & 262 & \textbf{ 237 } \\
invextr1\_.. & 3924 & 2542 & \textbf{ 948 } & 447 & 513 & \textbf{ 1244 } & 167 & 305 & \textbf{ 568 } \\
g7jac120s.. & 3025 & 3262 & \textbf{ 761 } & 104 & 165 & \textbf{ 236 } & 123 & 228 & \textbf{ 240 } \\
av41092 & 4944 & 2823 & \textbf{ 1317 } & 222 & 271 & \textbf{ 472 } & 161 & 212 & \textbf{ 233 }
\end{tabular}
\end{table}

\begin{table}[h]
\centering
\caption{An eigenvalue accuracy comparison between LAPACK (left in each column-triplet), ScaLAPACK (middle in each column-triplet) and StarNEig (right in each column-triplet) on a Skylake node (25 cores) and an Abisko node (24 cores/FPUs) when computing a Schur form from a Hessenberg form.}
\label{tab:schur_errors}
\begin{tabular}{r | c c c | c c c | c c c | c c c }
& \multicolumn{6}{c|}{Skylake node} & \multicolumn{6}{c}{Abisko node} \\
Matrix & \multicolumn{3}{c|}{$E_{\lambda, mean}$} & \multicolumn{3}{c|}{$E_{\lambda, max}$} & \multicolumn{3}{c|}{$E_{\lambda, mean}$} & \multicolumn{3}{c}{$E_{\lambda, max}$}\\
\hline
syn\_10000 & 78 & 44 & \textbf{ 62 } & 483 & 397 & \textbf{ 505 } & 78 & 64 & \textbf{ 59 } & 561 & 392 & \textbf{ 461 } \\
syn\_20000 & 89 & 84 & \textbf{ 92 } & 1032 & 1078 & \textbf{ 1488 } & 91 & 146 & \textbf{ 87 } & 992 & 1672 & \textbf{ 867 } \\
syn\_40000 & 116 & 179 & \textbf{ 181 } & 1687 & 2959 & \textbf{ 3462 } & 115 & 348 & \textbf{ 183 } & 1621 & 42734 & \textbf{ 1801 }
\end{tabular}
\end{table}

The first set of experiments demonstrates the performance of StarNEig compared to LAPACK and ScaLAPACK on a Skylake node (Table \ref{tab:keb_schur_sparce}) and an Abisko node (Table \ref{tab:abi_schur_sparce}).
The quantities $E_{\lambda, mean}$ and $E_{\lambda, max}$ are listed separately in Table \ref{tab:schur_errors} when applicable.
The results show that when compared against LAPACK, StarNEig is between 2.6 and 4.8 (geometric mean 3.7) times faster on a Skylake node and between 3.8 and 6.8 (geometric mean 5.1) times faster on an Abisko node.
When compared against ScaLAPACK, StarNEig is between 1.6 and 6.3 (geometric mean 2.6) times faster on a Skylake node and between 2.1 and 4.3 (geometric mean 3.2) times faster on an Abisko node.

\subsubsection{Comment regarding the residuals}

Note that although StarNEig is slightly less accurate than both LAPACK and ScaLAPACK, the differences are very small, in particular considering the observations made in Subsection \ref{subsec:deflation_cond}.
The differences are likely contributable to the way the algorithms compute the shifts since, under certain conditions, the LAPACK and ScaLAPACK algorithms recompute the shifts after an AED step in order to improve their quality.
The same functionality has been planned as an option for StarNEig.
More accurate shifts can lead to a smaller number of iterations and less roundoff error.
The fact that StarNEig is actually slightly slower than LAPACK in a single-core environment also supports this conclusion.

One of the reason why Subsection \ref{subsec:deflation_cond} is included in this paper is to counter the argument that StarNEig is faster because it somehow deflates eigenvalues earlier than LAPACK and ScaLAPACK even though all three libraries are configured to use the same deflation condition.
However, this is unlikely to be the case because, as shown in Subsection \ref{subsec:deflation_cond}, a less strict deflation condition leads to a slightly more accurate solution in most cases.
Furthermore, note that the residuals $R_{A}(\cdot,\cdot)$ and $R_{orth}(\cdot)$ actually start from zero and grow as more computations are performed.

\subsection{Single-node scalability}

\begin{figure}[h]
 \begin{center}
  \includegraphics[scale=0.65]{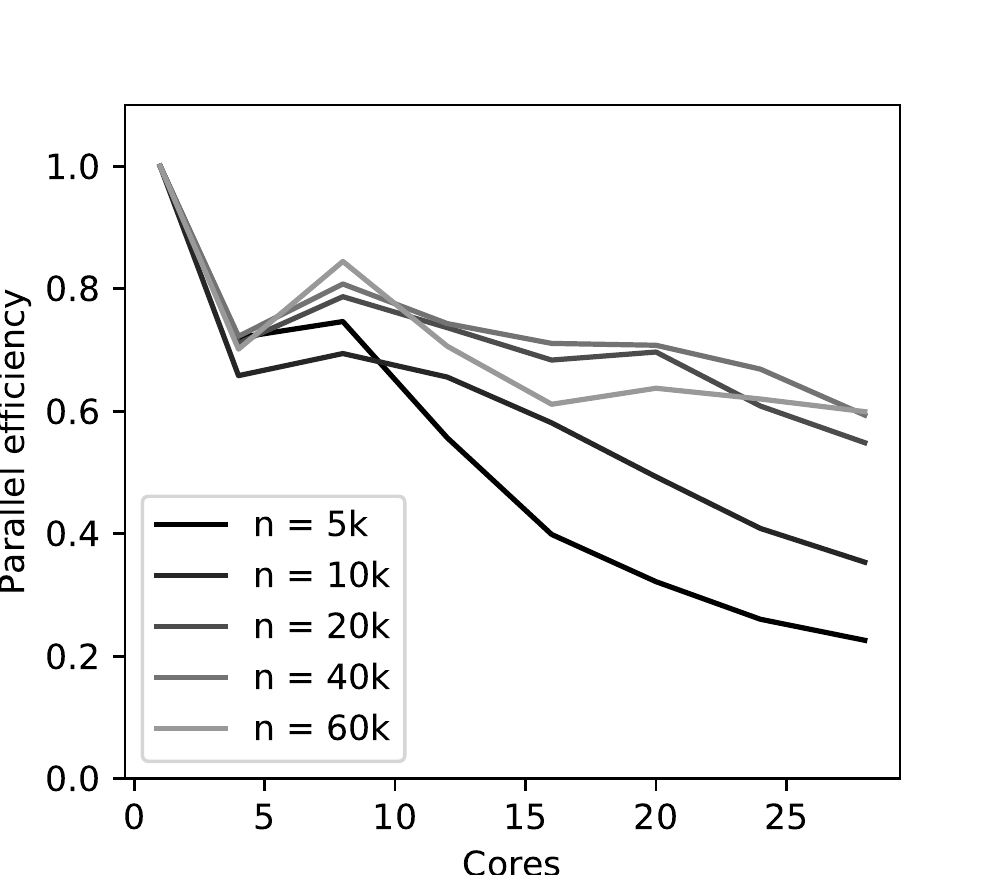}
  \includegraphics[scale=0.65]{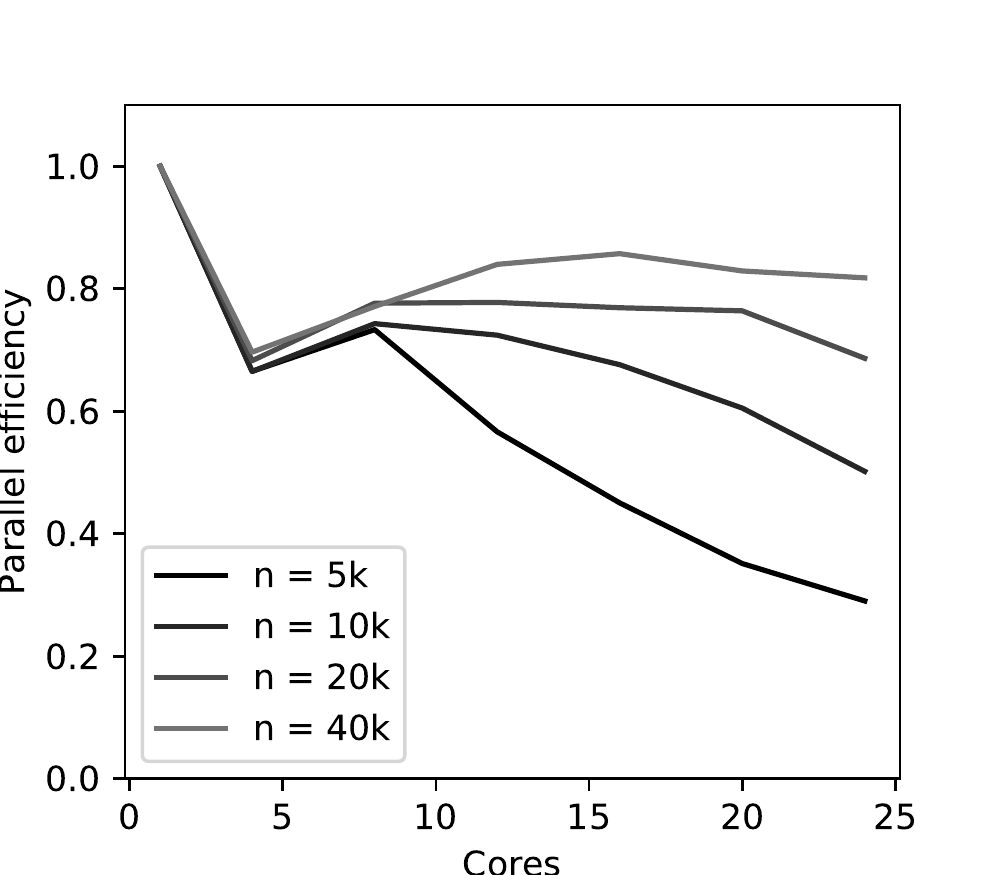}
 \end{center}
 \caption{
 Single-node parallel efficiency on a Skylake node (left) and an Abisko node (right) when computing a Schur form from a Hessenberg form hess\_$n$.
 }
 \label{fig:x_schur_sm}
\end{figure}

The second set of experiments demonstrates the scalability of StarNEig in single-node environments, see Fig. \ref{fig:x_schur_sm}.
The sharp dip at 4 cores in both graphs is due to the fact that StarNEig is configured to use only 3 StarPU workers (see Subsection \ref{subsection:starneig_conf}) and the relative loss in the available performance is at its highest in that instance.
If the number of StarPU workers was set to equal the number of available cores, then the dip at 4 cores would disappear but the performance would instead dip at 28 cores, since the thread that inserts the tasks into StarPU would share a core with a StarPU worker thread.
A similar effect could in principle occur at 1 core, and thus affect the shapes of the graphs, but this does not seem to occur in practice.
Note that both Skylake (base 2.60 GHz, turbo 3.70 GHz) and Abisko (base 2.60 GHz, turbo 3.20 GHz) nodes run at higher clock rate under a single-core load.
The exact Skylake boost table is proprietary information and has not been released by the manufacturer.
Therefore, neither graph is corrected for the changing clock rates.

\subsection{Multi-node performance comparison}

\begin{table}[h]
\centering
\caption{A comparison between ScaLAPACK (left in each column-pair) and StarNEig (right in each column-pair) on multiple Broadwell nodes when computing a Schur form from a Hessenberg form.
The missing ScaLAPACK results are due to the fact that the parallel file system used in Kebnekaise could not handle an I/O load where a large number of MPI processes are simultaneously reading from the same file.}
\label{tab:keb_schur_table}
\begin{tabular}{r | c c | c c | c c | c c | c c | c c}
\multicolumn{1}{c|}{syn\_$n$} & \multicolumn{2}{c|}{Cores} & \multicolumn{2}{c|}{Run time [s]} & \multicolumn{2}{c|}{$R_{A}(S,Q)$} & \multicolumn{2}{c|}{$R_{orth}(Q)$} & \multicolumn{2}{c|}{$E_{\lambda, mean}$} & \multicolumn{2}{c}{$E_{\lambda, max}$}\\ \hline
10\,000 & 121 & 112 & 27 & \textbf{ 10 } & 155 & \textbf{ 230 } & 100 & \textbf{ 147 } & 31 & \textbf{ 56 } & 398 & \textbf{ 386 } \\
20\,000 & 121 & 112 & 95 & \textbf{ 32 } & 234 & \textbf{ 334 } & 157 & \textbf{ 239 } & 61 & \textbf{ 93 } & 1128 & \textbf{ 1121 } \\
40\,000 & 121 & 112 & 451 & \textbf{ 184 } & 393 & \textbf{ 658 } & 280 & \textbf{ 421 } & 117 & \textbf{ 186 } & 1772 & \textbf{ 1848 } \\
60\,000 & 256 & 252 & 1004 & \textbf{ 396 } & 623 & \textbf{ 884 } & 402 & \textbf{ 556 } & 175 & \textbf{ 247 } & 1995 & \textbf{ 2257 } \\
80\,000 & 256 & 252 & 1679 & \textbf{ 735 } & 923 & \textbf{ 1099 } & 559 & \textbf{ 712 } & 250 & \textbf{ 313 } & 6049 & \textbf{ 3695 } \\
100\,000 & 256 & 252 & 3366 & \textbf{ 1420 } & 987 & \textbf{ 1377 } & 769 & \textbf{ 852 } & 336 & \textbf{ 380 } & 3964 & \textbf{ 4377 } \\
120\,000 & 484 & 448 & -- & \textbf{ 2179 } & -- & \textbf{ 1750 } & -- & \textbf{ 1114 } & -- & \textbf{ 495 } & -- & \textbf{ 6298 } \\
140\,000 & 484 & 448 & -- & \textbf{ 2351 } & -- & \textbf{ 2026 } & -- & \textbf{ 1238 } & -- & \textbf{ 564 } & -- & \textbf{ 7500 } \\
160\,000 & 484 & 448 & -- & \textbf{ 3360 } & -- & \textbf{ 2031 } & -- & \textbf{ 1397 } & -- & \textbf{ 628 } & -- & \textbf{ 9065 }
\end{tabular}
\end{table}

\begin{table}[h]
\centering
\caption{A comparison between ScaLAPACK (left in each column-pair) and StarNEig (right in each column-pair) on multiple Abisko nodes when computing a Schur form from a Hessenberg form.}
\label{tab:abi_schur_table}
\begin{tabular}{r | c c | c c | c c | c c | c c | c c}
\multicolumn{1}{c|}{syn\_$n$} & \multicolumn{2}{c|}{Cores/FPUs} & \multicolumn{2}{c|}{Run time [s]} & \multicolumn{2}{c|}{$R_{A}(S,Q)$} & \multicolumn{2}{c|}{$R_{orth}(Q)$} & \multicolumn{2}{c|}{$E_{\lambda, mean}$} & \multicolumn{2}{c}{$E_{\lambda, max}$}\\ \hline
10\,000 & 100 & 96 & 58 & \textbf{ 25 } & 155 & \textbf{ 223 } & 100 & \textbf{ 149 } & 32 & \textbf{ 57 } & 377 & \textbf{ 392 } \\
20\,000 & 100 & 96 & 206 & \textbf{ 91 } & 245 & \textbf{ 323 } & 164 & \textbf{ 234 } & 66 & \textbf{ 89 } & 893 & \textbf{ 1083 } \\
40\,000 & 100 & 96 & 1049 & \textbf{ 493 } & 385 & \textbf{ 605 } & 275 & \textbf{ 395 } & 115 & \textbf{ 173 } & 1929 & \textbf{ 1760 } \\
60\,000 & 225 & 216 & 2013 & \textbf{ 919 } & 645 & \textbf{ 849 } & 428 & \textbf{ 556 } & 186 & \textbf{ 250 } & 2805 & \textbf{ 3074 } \\
80\,000 & 225 & 216 & 4091 & \textbf{ 2061 } & 738 & \textbf{ 1158 } & 506 & \textbf{ 731 } & 211 & \textbf{ 301 } & 7439 & \textbf{ 3690 } \\
100\,000 & 225 & 216 & 8189 & \textbf{ 4288 } & 1081 & \textbf{ 1256 } & 748 & \textbf{ 835 } & 312 & \textbf{ 400 } & 3449 & \textbf{ 6688 }
\end{tabular}
\end{table}

The third set of experiments demonstrates the performance of StarNEig in comparison to ScaLAPACK on multiple Skylake and Abisko nodes, see Tables \ref{tab:keb_schur_table} and \ref{tab:abi_schur_table}.
Both ScaLAPACK and StarNEig were configured to use a square MPI process grid and ScaLAPACK was always given the advantage in terms of number of cores.
Technically it is possible to use exactly the same core count in both experiments.
Specifically, this would require us to solve the equation $\hat p_{scalapack}^2 = a \hat p_{starneig}^2$, where $\hat p_{scalapack}$ is the dimension of the square ScaLAPACK process grid, $\hat p_{starneig}$ is the dimension of the square StarNEig process grid, and $a$ is the number of cores per node used by StarNEig.
With 28 cores per node (Kebnekaise), the only choices of $a$ are 1, 2, 4, 9, 16 and 25.
Only in the last case of $a = 25$ are we using a reasonable fraction of the available cores.
With 24 cores/FPUs per node (Abisko), the only choices of $a$ are 1, 2, 4, 9, and 16.
Because we are using two different machines, the only viable course of action is to give ScaLAPACK the advantage in terms of cores and run both software packages using the intended optimal configuration.
The results show that when compared against ScaLAPACK, StarNEig is between 2.3 and 3.0 (geometric mean 2.5) times faster on the Broadwell nodes and between 1.9 and 2.3 (geometric mean 2.1) times faster on the Abisko nodes.

\subsection{Multi-node scalability}

\begin{figure}[h]
 \centering
 \includegraphics[scale=0.65]{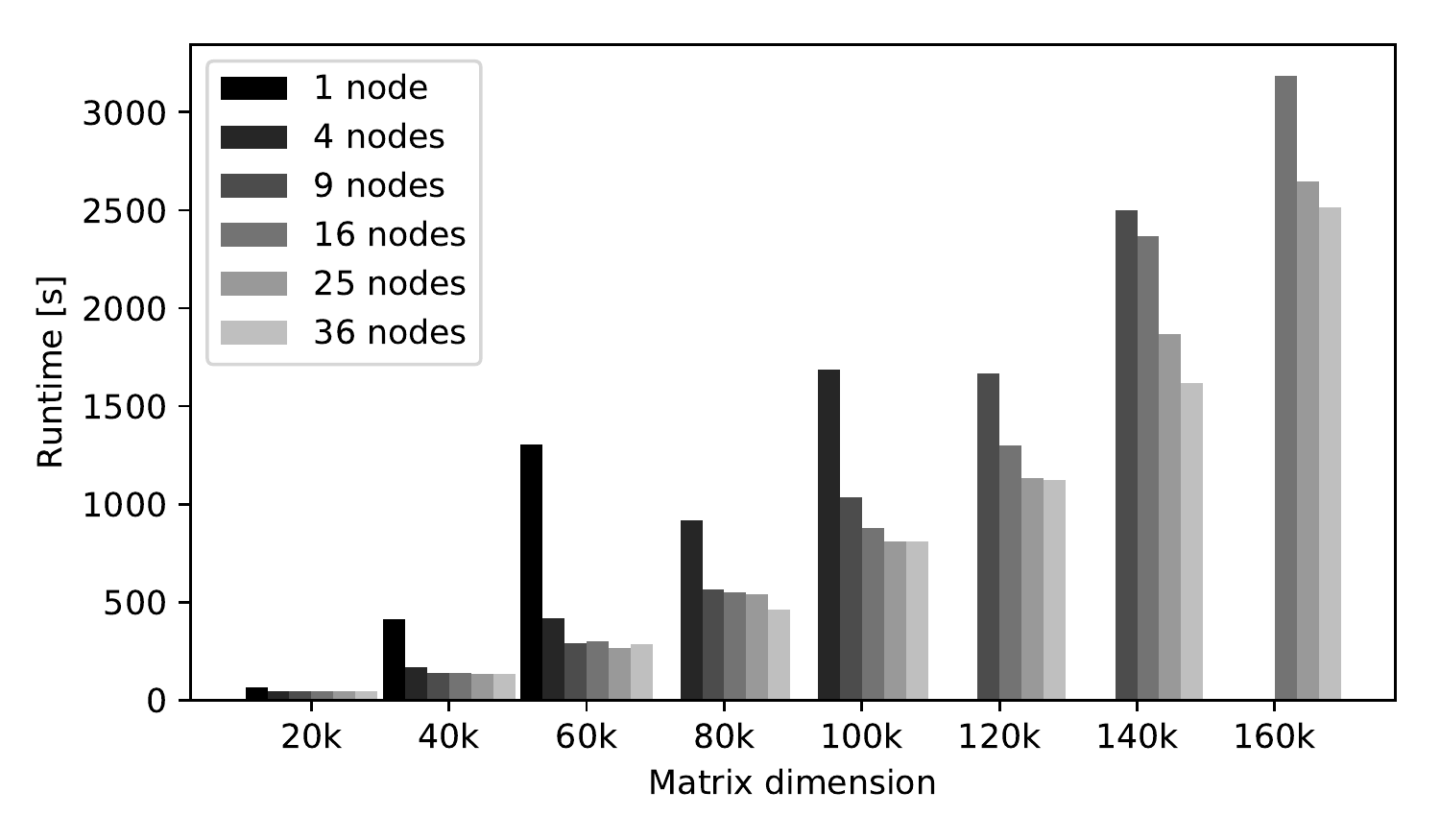}
 \caption{
 Multi-node scalability on Broadwell nodes when computing a Schur form from a Hessenberg form hess\_$n$.
 The missing data points are due to the fact that the matrices simply did not fit into the combined memory of the nodes. 
 }
 \label{fig:keb_schur_scale_starneig}
\end{figure}

\begin{figure}[h]
 \centering
 \includegraphics[scale=0.65]{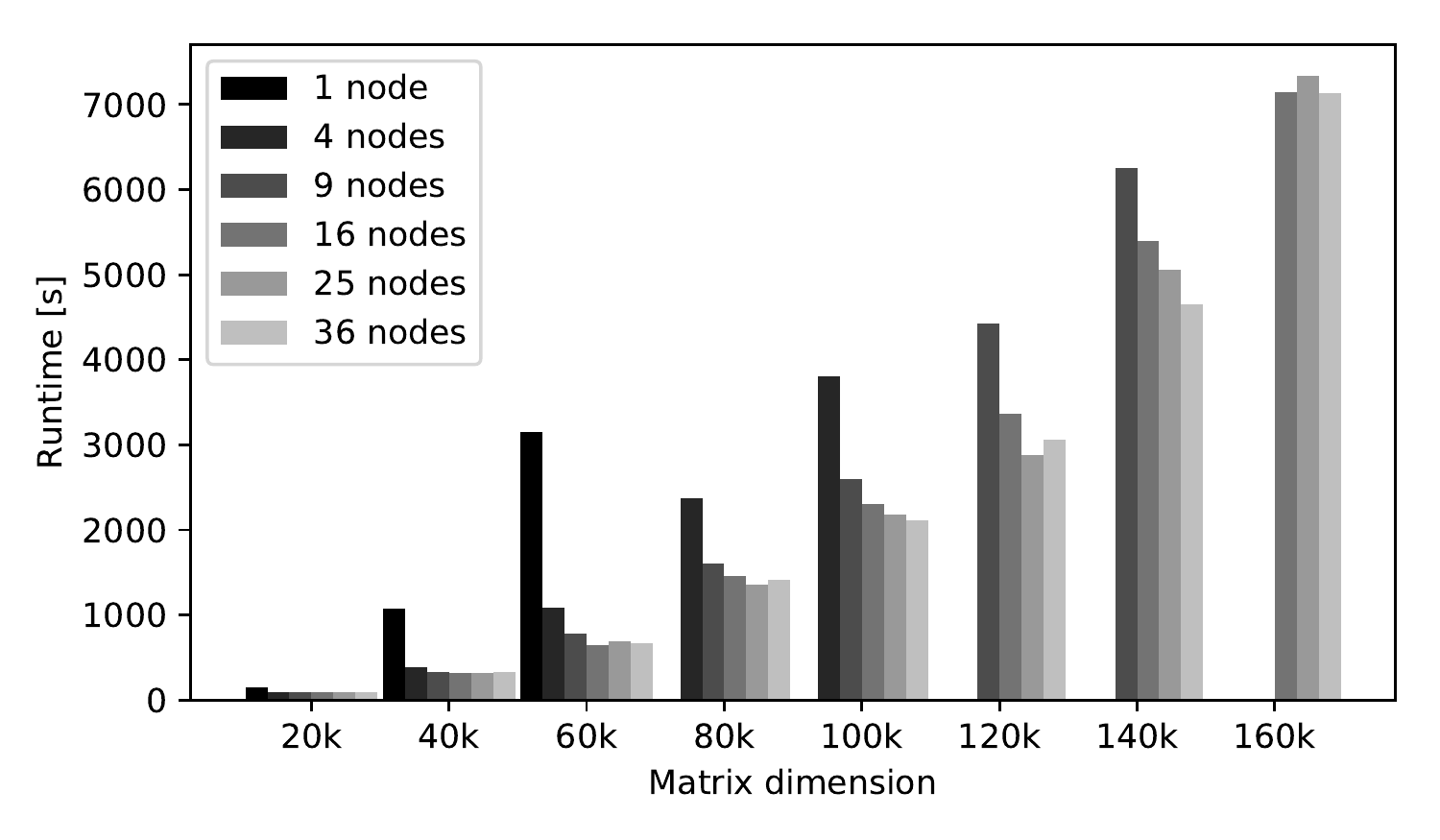}
 \caption{
 Multi-node scalability on Abisko nodes when computing a Schur form from a Hessenberg form hess\_$n$.
 The missing data points are due to the fact that the matrices simply did not fit into the combined memory of the nodes. 
 }
 \label{fig:abi_schur_scale_starneig}
\end{figure}

The fourth set of experiments demonstrates the scalability of StarNEig in multi-node environments.
The results can be seen in Figures \ref{fig:keb_schur_scale_starneig} and \ref{fig:abi_schur_scale_starneig}.
The tile size was fixed to 248-by-248 and the distributed block size to 1984-by-1984.
Overall, the scalability is very good when moving from a single node to four nodes but thereafter the benefits from the additional nodes begin to diminish.
Note that the behavior of the QR algorithm is sensitive to the properties of the matrix.
In particular, some matrices require only a few bulge chasing steps and thus converge very fast but do not offer that many opportunities for concurrency.

\subsection{GPU performance}

\begin{table}[h]
\centering
\caption{Run time (in seconds) on a V100 GPU node (28 cores + 0/1/2 GPUs) when computing a Schur form from a Hessenberg form.}
\label{tab:keb_schur_gpu}
\begin{tabular}{r | c | c | c }
Matrix & CPU only & CPU + GPU & CPU + 2 GPUs \\
\hline
syn\_10000 & 14 & 17 & 16 \\
syn\_20000 & 64 & 61 & 55 \\
syn\_40000 & 437 & 285 & 273 \\
\hline
g7jac020 & 3 & 4 & 4 \\
sinc15 & 9 & 10 & 10 \\
ex11 & 39 & 34 & 30 \\
ns3Da & 54 & 56 & 50 \\
TSOPF\_RS\_.. & 215 & 169 & 159 \\
invextr1\_.. & 373 & 160 & 143 \\
g7jac120s.. & 254 & 141 & 137 \\
av41092 & 454 & 305 & 316
\end{tabular}
\end{table}

The fifth set of experiments demonstrates StarNEig's GPU performance, see Table \ref{tab:keb_schur_gpu}.
The introduction of a single GPU leads to a speedup between 0.8 and 2.3 (geometric mean 1.2) and the introduction of two GPUs leads to a speedup between 0.8 and 2.6 (geometric mean 1.3).
The speedups are computed with respect to the CPU-only results.

The GPU support, and the multi-GPU support in particular, are still considered experimental features.
Also note that in case of the smaller matrices and with 28 cores, the execution time is likely to be bounded from below by the CPU-only {\tt push bulges} tasks.
Only {\tt right update} and {\tt left update} tasks are offloaded to the GPUs.
In other words, when the problem is relatively small and we have a large number of CPU cores at our disposal, the CPU cores can already execute all low-priority tasks at the same rate as the critical path is progressing. 
Therefore, the critical path is the deciding factor when it comes to the performance. 
Offloading low-priority tasks to the GPUs does not change this fact. 
If a smaller number of CPU cores were used, then the GPUs would have a larger impact.
The final thing to note is that as shown in \tcite{myllykoski_2020}, a two-sided matrix transformation algorithm can scale to multiple GPUs.

\section{Summary and conclusions}\label{sec:conclusions}

This paper presented a task-based multi-shift QR algorithm with AED and its StarPU-based implementation.
The task-based algorithm adopts previous algorithmic improvements, such as tightly-coupled multi-shifts and Aggressive Early Deflation (AED) from the latest LAPACK and ScaLAPACK algorithms.
On top of these, the task-based algorithm incorporates several new improvements:
\begin{enumerate}

 \item Elimination of several (global) synchronization points.
 This is a natural outcome of the task-based approach as the runtime system guarantees that tasks are executed in a sequentially consistent order.
 However, a naive task-based implementation can still contain several implicit (global) synchronization points.
 This was demonstrated in \tcite{D26} where an earlier version of the task-based algorithm was compared against LAPACK and ScaLAPACK.
 This early prototype code did not include most of the task insertion order and task priority optimizations discussed in Section \ref{sec:task_qr} and therefore scaled poorly even though it outperformed both LAPACK and ScaLAPACK.
  The task priorities are an important tool as the runtime system does not otherwise know which tasks are likely to stall the execution if not executed early enough.
 Furthermore, the task-based algorithm is designed to avoid situations where the main thread has to wait for a specific task to complete its execution.
 For example, as mentioned in Subsection \ref{subsec:eventdriven}, the \textbf{Bootstrap} and \textbf{Bulges} states are designed so that an action is triggered immediately for each newly discovered unreduced block.
 This is done because the main thread might get blocked when it is attempting to acquire the next section of the aftermath vector.
 Also, as explained in Subsection \ref{subsec:greedy_approach}, the task-based algorithm avoids acquiring the outcomes of most {\tt deflate} tasks by relying on a greedy approach.
 
 \item Dynamic merging of previously separate computational steps.
 As shown in Fig. \ref{fig:frames2}, an AED step can be executed concurrently with a bulge chasing step. 
 The task-based algorithm can even overlap two bulge chasing steps with each other as shown in the same figure.
 This cannot happen in ScaLAPACK (nor in LAPACK) because ScaLAPACK synchronizes before and after each AED step. 
 Furthermore, as shown in Fig. \ref{fig:frames}, the task-based algorithm allows the bulge chasing windows to operate almost completely independently of each other because the execution is not synchronized when the bulges are moved from one tile or a distributed block to another as done in ScaLAPACK.
 In the task-based algorithm, the theoretical maximum number of concurrent bulge chasing windows is the total number of worker threads whereas ScaLAPACK is limited to the square root of the total number of MPI processes.
 LAPACK is limited to just one concurrent bulge chasing window.
 
 \item Shortening and prioritization of the critical path. 
 Subsection \ref{sec:task_qr} discusses extensively how the task insertion order and the task priorities are used to accelerate the critical path and delay low-priority tasks.
 In particular, as discussed in Subsection \ref{subsec:shortening_path}, visualized in Fig. \ref{fig:chain_flow2} and partly demonstrated in Figures \ref{fig:frames} and \ref{fig:frames2}, the number of update tasks that need to be executed before the next AED step can begin has been reduced to close to a quarter of the number in ScaLAPACK.
 In ScaLAPACK, all bulge chasing related computations must be completed before the next AED step can begin.
 This means that most MPI processes are idle while a small subset of the MPI processes are performing an AED step.
 The task-based algorithm can use the otherwise idle cores to execute low-priority tasks.
 Furthermore, as discussed in Subsection \ref{subsec:lowest_prio}, update tasks that do not feed back to any unreduced block are given the lowest priority and are therefore delayed until computational resources start becoming idle.
 
 \item Concurrent processing of several unreduced diagonal blocks.
 Subsection \ref{subsec:eventdriven} discusses how the task-based algorithm keeps track and concurrently processes several unreduced blocks.
 As mentioned in the first item of this list, the task-based algorithm is designed so that the number of instances, where the main thread has to wait for a specific task to complete its execution, has been minimized.
 Added to this, as shown in Fig. \ref{fig:states}, the \textbf{AED reduce} and \textbf{AED deflate} states are designed so that the task insertion process can be interrupted.
 This means that the main thread can move on to processing the next unreduced blocks while the runtime system is processing tasks that are related to the current unreduced block. 
 ScaLAPACK cannot process multiple unreduced blocks concurrently.
 
 \item Adaptive approach for deciding when to perform a parallel AED.
 There are instances where it is not tentatively clear whether an AED step should be performed in parallel or not.
 As explained in Subsection \ref{subsec:adaptive_approach}, the task-based algorithm uses performance models to predict when a parallel AED is likely to be beneficial.
 This can reduce both the core idle time and the overall execution time.
 
 \item A common algorithm and implementation for both single-node and multi-node environments.
 The StarNEig implementation uses the same code base for both shared and distributed memory computations.
 Most of the work is offloaded to the StarPU runtime system but code base still contains several hand-written distributed memory specific sections.
 For example, the main thread must occasionally acquire the outcome of certain tasks.
 In these instances, the relevant data handles must be manually requested to be transferred to the node in question.
 Furthermore, all hand-written communications are asynchronous and thus do not block the main thread.
 
 \item GPU acceleration. 
 The StarPU runtime system can offload computations to GPUs.
 In particular, the task-based algorithm allows StarPU to offload most BLAS-3 operations to GPUs.
 This could theoretically lead to huge performance improvements as GPUs are exceptionally fast with BLAS-3 operations.

\end{enumerate}

The implementation was demonstrated to be many times faster than multi-threaded LAPACK and ScaLAPACK in both single-node and multi-node environments on Intel and AMD based machines.
Furthermore, the experimental GPU acceleration was demonstrated to be functional in single-GPU configuration.

\section*{Acknowledgments} 
\label{sec:acknowledgements}

\paragraph{StarNEig authors}
StarNEig has been developed by the author (Hessenberg reduction, Schur reduction, eigenvalue reordering, integration and testing), Carl Christian Kjelgaard Mikkelsen (generalized eigenvectors and theoretical contributions to the robust computation of eigenvectors), Angelika Schwarz (standard eigenvectors), Lars Karlsson (miscellaneous contributions to the library and theoretical contributions to the robust computation of eigenvectors), and Bo K{\aa}gstr{\"o}m (coordinator and scientific director of the entire NLAFET project).

\paragraph{Other acknowledgments}
The author would like to separately extend his gratitude to Carl Christian Kjelgaard Mikkelsen and Birgitte Bryds{\o} for their valuable comments and suggestions during the preparation of this paper.
The author also acknowledges the work of Mahmoud Eljammaly and Bj\"{o}rn Adlerborn during the NLAFET project.
The author thanks the High Performance Computing Center North (HPC2N) at Ume{\aa} University, for providing computational resources and valuable support during test and performance runs, and the StarPU team at INRIA, for rapidly providing good answers to several questions related to StarPU.
Finally, the author thanks the anonymous reviewers for their valuable feedback.

\bibliographystyle{ACM-Reference-Format}
\citestyle{acmauthoryear}
\bibliography{references}

\end{document}